\documentclass[twocolumn]{aastex63}
\usepackage{natbib}
\usepackage{multirow}
\graphicspath{{./}{figures/}}
\usepackage{color,subfigure,lineno} 
\shorttitle{VODKA: VLBA Survey}
\shortauthors{Chen et al.}
\begin{document}

\title{Varstrometry for Off-nucleus and Dual Sub-kpc AGN (VODKA): Very Long Baseline Array Searches for Dual or Off-nucleus Quasars and Small-scale Jets}

\correspondingauthor{Yu-Ching Chen}
\email{ycchen@jhu.edu}

\author[0000-0002-9932-1298]{Yu-Ching Chen}
\affiliation{Department of Astronomy, University of Illinois at Urbana-Champaign, Urbana, IL 61801, USA}
\affiliation{National Center for Supercomputing Applications, University of Illinois at Urbana-Champaign, Urbana, IL 61801, USA}
\affiliation{Center for AstroPhysical Surveys, National Center for Supercomputing Applications, Urbana, IL, 61801, USA}

\author[0000-0003-0049-5210]{Xin Liu}
\affiliation{Department of Astronomy, University of Illinois at Urbana-Champaign, Urbana, IL 61801, USA}
\affiliation{National Center for Supercomputing Applications, University of Illinois at Urbana-Champaign, Urbana, IL 61801, USA}
%\affiliation{Center for AstroPhysical Surveys, National Center for Supercomputing Applications, Urbana, IL, 61801, USA}
\affiliation{Center for Artificial Intelligence Innovation, University of Illinois at Urbana-Champaign, 1205 West Clark Street, Urbana, IL 61801, USA}

\author{Joseph Lazio}
\affiliation{Jet Propulsion Laboratory, California Institute of Technology, Pasadena, CA 91109, USA}

\author[0000-0003-1317-8847]{Peter Breiding}
\affiliation{Department of Physics and Astronomy, Johns Hopkins University, Baltimore, MD 21218, USA}

\author{Sarah Burke-Spolaor}
\affiliation{Department of Physics and Astronomy, West Virginia University, Morgantown, WV 26506, USA}
\affiliation{Center for Gravitational Waves and Cosmology, West Virginia University, Chestnut Ridge Research Building, Morgantown, WV 26505, USA}

\author[0000-0003-4250-4437]{Hsiang-Chih Hwang}
\affiliation{School of Natural Sciences, Institute for Advanced Study, Princeton, NJ 08540, USA}

\author[0000-0003-1659-7035]{Yue Shen}
%\altaffiliation{Alfred P. Sloan Research Fellow}
\affiliation{Department of Astronomy, University of Illinois at Urbana-Champaign, Urbana, IL 61801, USA}
\affiliation{National Center for Supercomputing Applications, University of Illinois at Urbana-Champaign, Urbana, IL 61801, USA}

\author[0000-0001-6100-6869]{Nadia L. Zakamska}
\affiliation{Department of Physics and Astronomy, Johns Hopkins University, Baltimore, MD 21218, USA}

%\author[0000-0001-6100-6869]{Qian Yang}
%\affiliation{Department of Astronomy, University of Illinois at Urbana-Champaign, Urbana, IL 61801, USA}

%\author[0000-0002-0311-2812]{Jennifer I. Li}
%\affiliation{Department of Astronomy, University of Illinois at Urbana-Champaign, Urbana, IL 61801, USA}

\begin{abstract}
Dual and off-nucleus active supermassive black holes are expected to be common in the hierarchical structure formation paradigm, but their identification at parsec scales remains a challenge due to strict angular resolution requirements.
We conduct a systematic study using the Very Long Baseline Array (VLBA) to examine 23 radio-bright candidate dual and off-nucleus quasars. The targets are selected by a novel astrometric technique (``varstrometry'') from Gaia, aiming to identify dual or off-nucleus quasars at (sub)kilo-parsec scales.
Among these quasars, 8 exhibit either multiple radio components or significant ($>3\sigma$) positional offsets between the VLBA and Gaia positions. The radio emission from the three candidates which exhibit multiple radio components is likely to originate from small-scale jets based on their morphology. Among the remaining five candidates with significant VLBA-Gaia offsets, three are identified as potential dual quasars at parsec scales, one is likely attributed to small-scale jets, and the origin of the last candidate remains unclear. 
We explore alternative explanations for the observed VLBA-Gaia offsets. We find no evidence for optical jets at kilo-parsec scales, nor any contamination to Gaia astrometric noise from the host galaxy; misaligned coordinate systems are unlikely to account for our offsets. 
%Comparisons with samples of single normal quasars from the literature suggest that varstrometry can select a higher fraction of quasars with significant VLBA-Gaia offsets. 
Our study highlights the promise of the varstrometry technique in discovering candidate dual or off-nucleus quasars and emphasizes the need for further confirmation and investigation to validate and understand these intriguing candidates.

\end{abstract}

\keywords{accretion, accretion disks --- black hole physics --- galaxies: active --- galaxies: jets -- quasars: general --- radio continuum: galaxies -- reference systems -- surveys}

\section{Introduction}\label{sec:introduction}

%Since most massive galaxies are believed to harbor a central supermassive black hole (SMBH), galaxy mergers result in the formation of binary SMBHs \citep{begelman80}. Despite the success of the merger scenario in explaining much of the observed phenomenology of AGN statistics (such as luminosity function and clustering properties \citep{kauffmann00,volonteri03,hopkins08,shen09}), direct observational evidence for binary SMBHs remains scarce. Only one confirmed case is known at $\ll$1 kpc \citep{rodriguez06}, in stark contrast to theoretical expectations.  

%Since most massive galaxies are believed to harbor a central supermassive black hole \citep[SMBH;][]{kormendy95,KormendyHo2013}, binary SMBHs are expected to form from galaxy mergers \citep{begelman80}. Binary SMBHs are important for the studies of gravitational waves, galaxy evolution, and black hole astrophysics \citep{DeRosa2020,Bogdanovic2021}. Despite the success of the merger scenario in explaining much of the observed phenomenology of the quasar statistics \citep[such as the luminosity function and clustering properties; e.g.,][] {kauffmann00,volonteri03,hopkins08,shen09} and that numerical simulations have modeled the dynamical evolution of binary SMBHs down to tens-of-pc scales \citep[e.g.,][]{Fiacconi2013,Mayer2013,SouzaLima2017,Tamburello2017}, direct observational evidence for binary SMBHs and sub-kpc inspirally SMBHs remains scarce. Only one confirmed case is known at $\ll$1 kpc \citep{Rodriguez2006}, in stark contrast to naive theoretical expectations.  

Most massive galaxies are believed to harbor a central supermassive black hole \citep[SMBH;][]{kormendy95,KormendyHo2013}, and as a result, binary SMBHs are expected to form from galaxy mergers \citep{begelman80}. Binary SMBHs are of utmost importance to the studies of gravitational waves, black hole astrophysics, and galaxy evolution. Despite successful numerical simulations that have modeled the dynamical evolution of binary SMBHs down to parsec (pc) scales \citep{Fiacconi2013,Mayer2013,SouzaLima2017,Tamburello2017}, direct observational evidence for binary SMBHs and sub-pc inspiraling SMBHs remains scarce. 
%To date, only one confirmed case of a binary SMBH is known at pc scales \citep{Rodriguez2006}.
To date, the only known confirmed compact binary SMBH at pc scales is 0402+379, a flat-spectrum compact double radio source at $z=0.055$ with a projected separation of 7 pc, which was discovered serendipitously \citep{Rodriguez2006,Bansal2017}.
%This is in stark contrast to naive theoretical expectations.

%Another outstanding question is theory suggests that the orbital decay of binary SMBHs may slow down or even stall at $\sim$1--10 pc scales \citep{milosavljevic01,yu02}, or the barrier may be overcome by interactions with gas, stars, and/or a third BH \citep{gould00,hoffman07,Kelley2017}. The ``final-pc'' problem---that is, we don't know how efficiently SMBH binaries evolve through parsec-scale separations---poses the largest uncertainty in the expected intensity of low-frequency gravitational waves \citep{Arzoumanian2018a}. The observed rate of inspiraling SMBHs at sub-kpc scales and compact binaries at $\sim$pc scales thus can provide measurements that feed directly into final-pc physics, and the future interpretation of gravitational-wave detections.
A binary SMBH's orbital decay may slow down or halt at pc scales, a process commonly referred to as the ``final pc problem", because close encounters with stars are insufficient to remove orbital energy \citep{begelman80,Milosavljevic2001,Yu2002}. 
%and the gravitational emissions are still too weak at pc separations. 
While interactions with gas, triaxial galaxy structure, or a third SMBH may overcome this barrier \citep{Gould2000,Hoffman07,Kelley2017}, how efficiently binary SMBHs evolve from kpc to pc-scale separations remains uncertain. %due to limited observational evidence.
This uncertainty creates ambiguity in predicting the intensity of low-frequency gravitational waves by current facilities such as pulsar timing array \citep{Arzoumanian2018a}. To address this issue, we need to observe the rate of dual SMBHs at sub-kpc scales and compact binaries at $\sim$pc scales. These observations could provide direct measurements and insights into final-pc physics and the interpretation of future low-frequency gravitational-wave detections \citep{Bogdanovic2022}.

%The orbital decay of a binary SMBH may significantly slow down or even stall at $\sim$pc scales, known as the ``final-parsec" problem \citep{begelman80,Milosavljevic2001,Yu2002}. While the barrier may be overcome by interactions with gas, stars, and/or a third SMBH \citep{Gould2000,Hoffman07,Kelley2017}, observationally it is still  largely unknown how efficiently binary SMBHs evolve through the sub-kpc to pc-scale separations \citep{Shen2022}. This poses a significant uncertainty in the expected intensity of low-frequency gravitational waves \citep{Arzoumanian2018a}. The observed rate of inspiraling SMBHs at sub-kpc scales and compact binaries at $\sim$pc scales thus can provide measurements that feed directly into the final-pc physics and the interpretation of future low-frequency gravitational-wave detections \citep{Bogdanovic2021}.

Quasars are luminous sources of electromagnetic radiation powered by the accretion of matter onto the SMBH in active galactic nuclei (AGNs). 
%They have proven to be a valuable tool for identifying and studying dual and binary SMBHs \citep{Hennawi06,Liu11,Shen2023}. 
Following the eventual merger of a binary SMBH, the coalesced SMBH may receive a kick due to the anisotropic gravitational wave radiation because of momentum conservation \citep{Baker2006,Campanelli2007}. If the recoiled SMBH accretes from the ambient gas, it may emit radiation and be observed as an off-nucleus quasar \citep{Loeb07}. The observed statistics of off-nucleus quasars can be used to investigate the fueling processes of quasars in galaxy mergers, as well as the population of recoiling SMBHs that result from binary SMBH coalescence \citep{Blecha2016,Tremmel2018}.

The identification of sub-kpc dual/off-nucleus quasars has been challenging due to the strict spatial resolution required for their detection \citep{ChenYC2022}. Although an increasing number of sub-kpc dual AGN are being discovered serendipitously \citep{Woo14,Muller-Sanchez15,Goulding19,Koss23}, the systematic identification of sub-kpc dual/off-nucleus quasars has remained inefficient. 
%The only known confirmed compact binary SMBH is B0402+379, a flat-spectrum compact double radio source at $z=0.055$ with a projected separation of 7 pc, which was discovered serendipitously \citep{Rodriguez2006}.

We have developed a novel astrometric approach called ``varstrometry'' to identify sub-kpc unresolved dual/off-nucleus/lensed quasars throughout the entire sky (\citealt{HwangShen2020,ShenHwang2019}; see also \citealt{Williams1995,LiuY_2015,Springer2021,Mannucci2022}). With Gaia's coverage of the full sky and its depth reaching down to Gaia $G{\sim}21$ mag \citep[encompassing ${\sim}10^6$ AGN,][]{Carnerero2022}, varstrometry enables a systematic exploration of unresolved dual/off-nucleus AGNs in the poorly studied range at projected physical separations of $\sim$10--1000 pc. The technique's effectiveness has been established using Gaia DR2 data on a sample of
variable pre-main sequence stars, which have known unresolved close companions \citep{HwangShen2020}.
Hubble Space Telescope (HST) high-resolution dual-band optical imaging discovered dozens of quasars with kpc-scale double cores from varstrometry-selected targets \citep{ChenYC2022}, some of which were subsequently verified as kpc-scale dual or lensed quasars through extensive follow-up observations \citep{Shen2021,ChenYC2023}. 
Nonetheless, resolving sources at pc spatial scales remains challenging due to the angular resolution restriction of most existing facilities.  
Long-baseline radio or optical/infrared interferometers are a few instruments capable of directly imaging binary SMBHs with an angular resolution of a few milliarcseconds \citep[e.g.,][]{Rodriguez2006,Burke2011,Wrobel2014,GRAVITY,Sturm2018,Liu2018,EHT-M87,EHT-SgrA}. 

%In addition, the discovery of B0402+379 by the Very Long Baseline Array (VLBA) proves its capability to identify binary SMBHs at pc scales.
While radio interferometers are excellent tools to resolve sources that are not resolvable as binaries in optical surveys, finding two radio-loud sources in the same pair is statistically difficult if radio loudness is not affected by merging, because only $\sim$20\% of quasars are radio-loud \citep{Kellermann2016}. Previous systematic searches for binary SMBHs with with Very Long Baseline Interferometer (VLBI) have found a few candidates \citep{Burke2011,Tremblay2016,Breiding2021}. However, only one confirmed binary SMBH at pc scales, 0402+379, is known to date \citep{Rodriguez2006}. Gaia astrometry in comparison with VLBI astrometry is useful for studying jet systems of quasars \citep{Kovalev2017,Petrov2017,Makarov2017,Plavin2019,Petrov2019} or discovering binary/off-nucleus quasars \citep{Skipper2018,Breiding2021}. A recent study using varstrometry with e-MERLIN revealed four targets with Gaia-radio offsets of $\sim$ 9–60 mas \citep{Wang2023}, highlighting the potential of varstrometry to uncover candidate dual/off-nucleus quasars. In addition to binary SMBHs, VLBI was also used to search for strong gravitational lenses on milliarcsec scales \citep{Casadio2021}, which might be detected by varstrometry.

\begin{figure*}
    \centering

    \includegraphics[width=0.98\textwidth]{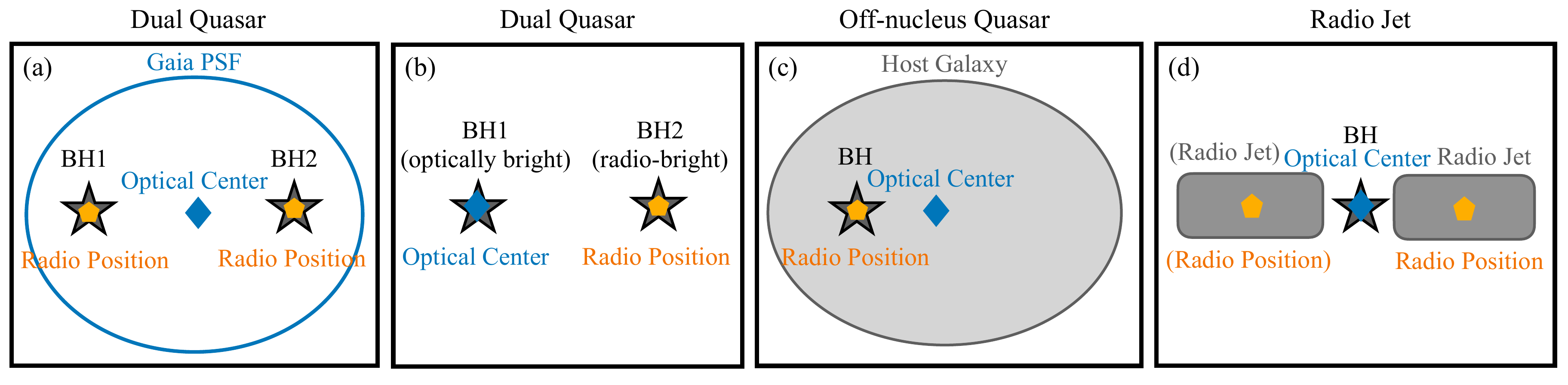}
    
    \caption{Illustration of different physical origins that can cause VLBA-Gaia offsets. The scales of offsets shown here are usually $<$ 0\farcs2, which is the Gaia's point spread function (blue ellipse). The optical position from Gaia is an unresolved weighted center (blue diamond), while the radio emission (yellow hexagon) is resolved with VLBA's mas angular resolution.
    (a): Dual quasar: The optical emission of two quasars is blended in Gaia resulting in a weighted optical center, while the radio emission is centered at individual quasar because of VLBA's high angular resolution.
    (b): Dual quasar: If only one quasar is radio-bright, the binary quasar show an optical-radio offset.
    (c): Off-nucleus quasar: The optical center is a weighted position of an off-nucleus quasar and a host galaxy, while the radio position is centered at the quasar's position. 
    (d): Radio jet from a single quasar: The optical center is centered at the quasar's position, while the radio position is shifted toward the jet direction.}
    \label{fig:cartoon}
\end{figure*}

In this study, we present observations of 23 radio-bright quasars at redshifts $0.5{<}z{<}2.5$, obtained with VLBA at X, C, and/or Ku-bands. These quasars were selected using Gaia data and the varstrometry technique \citep{ShenHwang2019,HwangShen2020}. While our earlier varstrometry papers \citep{Shen2021,ChenYC2022} focused on dual quasars at kpc scales, varstrometry can in principle probe $\sim$10-100 parsec scales. We discover six candidates with significant Gaia-VLBA offsets and three candidates with multiple radio cores at physical scales of $\sim$10--100 pc, which could be interpreted as binary quasars, off-nucleus quasars, or quasars with small-scale jets (see Figure \ref{fig:cartoon}). In this paper, we describe our methodology and sample selection in \S\ref{sec:sample}, and our VLBA observations, data reduction, and analysis in \S\ref{sec:observations}. Our main results are presented in \S\ref{sec:results}, and we discuss their implications in \S\ref{sec:discussions}. We summarize our findings and discuss future prospects in \S\ref{sec:conclusions}. Throughout this paper, all physical separations refer to projected separations. We adopt a flat $\Lambda$CDM cosmology with $\Omega_\Lambda=0.7$, $\Omega_m=0.3$, and $H_0=70\,{\rm km\,s^{-1}Mpc^{-1}}$.

\section{Sample selection}\label{sec:sample}

\begin{figure*}
\centering
 \includegraphics[width=0.47\textwidth]{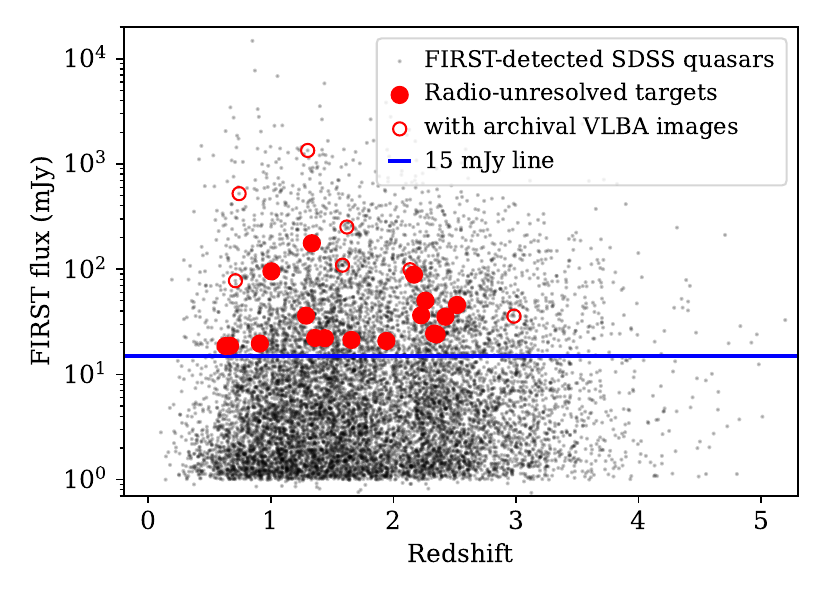}
 \includegraphics[width=0.47\textwidth]{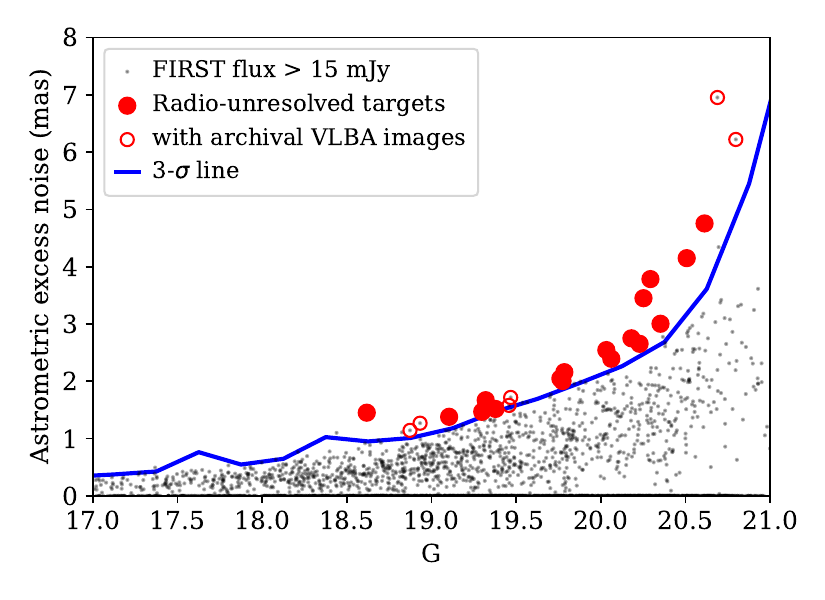}
 \caption{
 %\textbf{Sample selection, target properties, and a proof of concept.} 
 %\textbf{\color{red} Please enlarge the VLBA-related symbols so that it's easier to see. Remove the ``VLA resolved w/o archival VLBA images'' symbols because they are not included in the present study and do not need to be emphasized on the plot. Replace ``sigma'' with ``$\sigma$''. Make all the label texts bigger and lines thicker, if possible.} \textbf{\color{blue} YCC: Done.}
 Sample selection and the target properties. We selected the targets based on their peak flux densities ($>$15 mJy) in FIRST and astrometric excess noise ${>}3\sigma$ at a given Gaia magnitude.
 {\it Left}: 
 FIRST flux density versus redshift for the final analysis sample of 23 targets (marked in red) compared to all FIRST-detected sources in the parent quasar sample (marked in black), with a blue horizontal line indicating the 15 mJy threshold.
 %The 23 targets having significant astrometric excess noise (${>}3\sigma$ among all SDSS quasars at a given Gaia magnitude) are marked in red. 
 {\it Right}: 
  Astrometric excess noise versus Gaia magnitude for the final analysis sample of 23 targets (marked in red) compared to all radio-bright ($>$15 mJy) SDSS quasars (marked in balck), with a blue curve indicating the ${>}3\sigma$ threshold of astrometric excess noise at a given Gaia magnitude, based on all SDSS quasars.
 }
 %The 23 targets  are marked in red.   
%{\it Right}: A high astrometric noise AGN at $z$=2.983 selected from Gaia data with archival VLBA astrometry observations\citep{LiuHwang2019}. The source is unresolved by VLBA, but the radio centroid is significantly offset (at the 6.7$\sigma$ confidence level) from the optical position by 4.0 mas (corresponding to $\sim$30 pc at the redshift of the target) at PA$\sim$70$^{\circ}$ E of N. We suggest that this object is a binary AGN, composed of one radio-quiet source not seen in the VLBA data and another radio-brighter source seen in the image. The optical emission is contributed by both AGN, but the radio-quiet is likely brighter to produce the observed offset between the radio and the optical centroid. This candidate is a major success of our new Gaia-based selection technique. 
 
 \label{fig:target_selection}
\end{figure*}

%After carefully rejecting interlopers (e.g., extended hosts, stars with high proper motion and parallax), we have selected a sample of candidate sub-kpc dual/off-nuclear/lensed AGN from Gaia DR2 for further follow up. Figure \ref{fig:target_selection} shows our sample selection and target properties.
%\textcolor{red}{YCC:  We should chat about the selection in more details. I didn't edit this part.}
%{\color{blue} XL:Sounds good! let's definitely chat about this. I can take care of the writing of this subsection, but would need your help filling in the two tables listing basic properties of the new and archival targets.}

The selection of dual/offset quasar candidates relies on a novel astrometric technique, varstrometry, which exploits the centroid jitters resulting from a quasar's brightness variability \citep{LiuY_2015,ShenHwang2019,HwangShen2020}.
In the case of a dual quasar, these jitters arise from the non-coherent variability of each nucleus, causing the centroid of the total flux to shift. For an off-nucleus quasar, these jitters result from its brightness changing relative to a non-variable host galaxy, causing the centroid of the total flux to shift away from the center of the host galaxy as the flux of the quasar brightens. These astrometric jitter signals, induced by variability, can be detected by Gaia as ``excess astrometric noise" beyond the best-fit astrometric solution, or even as an anomaly in parallaxes and proper motions \citep{ShenHwang2019,HwangShen2020,ChenYC2022}. 

We focus on optically unobscured, broad-line quasars, for which the varstrometry technique is applicable.  Our study began with a sample of 356,850 spectroscopically confirmed quasars from Sloan Digital Sky Survey (SDSS) Data Release 14 \citep[DR14,][]{Paris2018}  that have single matches within 2\arcsec\ radius in Gaia Data Release 2 (DR2). Among these, we identified 3,735 objects with peak flux densities ${>}15$ mJy in Faint Images of the Radio Sky at Twenty-Centimeters \citep[FIRST,][]{Becker1995}. 
We then selected sources with significant Gaia DR2 astrometric excess noise ({\tt astrometric\_excess\_noise} ${>}3\sigma$ at any given Gaia magnitude, where $\sigma$ is the standard deviation of AEN as a function of G) to obtain 27 targets (Figure \ref{fig:target_selection}). AEN selection also helps to avoid strongly variable sources such as blazars (Fig. 12 in \citet{HwangShen2020}). We removed interlopers caused by extended host galaxies in PanSTARRS images \citep{Chambers2016} and star superpositions by inspecting SDSS optical spectra that enclose all fluxes within a 2\arcsec--3\arcsec\ diameter circle.
%In addition to FIRST, we also check existing VLA archival data to confirm that our targets remain compact on sub-arcsec scales and to improve the position information.

Among the 27 candidates, we found that seven have archival VLBA C-band (5 GHz) observations \citep{Taylor2005,Helmboldt2007}. We removed 1 archival target that lacks an accurate VLBA position from \citet{Petrov2011}. Of the remaining 20 candidates without archival VLBA images, three were excluded with FIRST-resolved jets, and then we conducted new VLBA observations for the rest 17 candidates. Our final sample consists of 23 targets that were observed in our new VLBA observations (18 targets) or have archival VLBA observations with accurate positions (6 targets). 
Only our target selection was based on Gaia DR2 AEN and magnitude (Figure \ref{fig:target_selection}); all the analyses in this paper are based on the new Gaia DR3 data. Gaia DR3 has notably improved its AEN measurements. The AENs decrease significantly from DR2 to DR3 for many of our targets (Table \ref{tab:target_info}), especially those without multiple radio components and offsets, which would no longer pass the 3$\sigma$ selection criteria if Gaia DR3 AEN is used. Properties of the 23 targets are listed in Table \ref{tab:target_info} including coordinate, redshift, magnitude, Gaia astrometric excess noise, Gaia flux variability, VLBA observed band, and FIRST flux density.

%The final sample consists 18 quasars including 17 new targets, which are point sources in FIRST and have no archival VLBA observations, and a VLBA-Gaia offset target, which only has archival VLBA C band observations.  \autoref{tab:target_info} lists basic properties of the 18 targets with new VLBA observations.
%and the \textcolor{red}{7 (need to double check)} with archival VLBA data, respectively.

%\startlongtable
\begin{deluxetable*}{lcccccccccc}
 \tablecaption{Basic properties of the final sample of 23 targets.
 \label{tab:target_info}}
 \tablehead{\colhead{Name} & \colhead{R.A.$_{\rm Gaia}$} & \colhead{Decl.$_{\rm Gaia}$} & \colhead{Redshift}  & \colhead{$r$} & \colhead{$G$} & \colhead{AEN} & \colhead{AEN$_{\rm DR2}$} & \colhead{Std.(Flux)}  & \colhead{Band} & \colhead{$S_{{\rm FIRST}}$} \\ 
  & \colhead{(J2000)} & \colhead{(J2000)}  &  & \colhead{(mag)} & \colhead{(mag)} & \colhead{(mas)} & \colhead{(mas)} & (e$^{-}$ s$^{-1}$) & & \colhead{(mJy)}  \\ 
 \colhead{(1)} & \colhead{(2)} & \colhead{(3)} & \colhead{(4)} & \colhead{(5)} & \colhead{(6)} & \colhead{(7)} & \colhead{(8)} & 
 \colhead{(9)} & \colhead{(10)} 
 }
 \startdata
 J0035$-$0911 & 00:35:53.0323 & $-$09:11:50.150 & 1.002 & 19.29 & 19.39 & 1.390 & 1.517 & 36.3  & X & 101.63  \\
 J0108$-$0400 & 01:08:28.1867 & $-$04:00:44.993 & 1.942 & 20.41 & 20.04 & 1.135 & 2.393 & 15.5 & C, X, Ku & 21.44 \\
 %J0202$-$0145\tablenotemark{b} & 02:02:27.3178 & -01:45:22.863 & 1.119 & 18.29 & 18.44 & 1.102 & 32.9 & \nodata & 30.77\\
 J0238+0123 & 02:38:22.4507 & +01:23:53.212 & 1.441 & 18.73 & 19.36 & 1.112 & 1.670 & 35.8 & C, X & 21.60 \\
 J0740+2852 & 07:40:33.5438 & +28:52:47.246 & 0.711 & 19.28 & 19.38 & 0.976 & 1.718 & 50.7 & C\tablenotemark{a} & 78.48\\
 J0749+2255 & 07:49:22.9664 & +22:55:11.770 & 2.166 & 19.15 & 18.53 & 1.482 & 1.452 & 160.1 & C, X, Ku & 91.95 \\
 J0840+4439 & 08:40:51.2889 & +44:39:59.291 & 0.769 & 20.08 & 20.14 & 1.967 & 2.654 & 22.2 & X & 180.26 \\
 %J0935+3933\tablenotemark{b} & 09:35:12.0731 & +39:33:00.919 & 1.003 & 21.27 & 20.70 & 4.342 & 19.3 & \nodata & 110.46 \\ 
 J1044$+$2959 & 10:44:06.3426 & +29:59:01.003 & 2.981 & 19.00 & 18.88 & 1.614 & 1.269 & 40.5 & C\tablenotemark{a}, X, Ku & 37.24 \\
 J1046+4827 & 10:46:57.1588 & +48:27:23.793 & 1.286 & 19.40 & 19.23 & 0.716 & 1.470 & 22.9 & X & 37.54 \\
 J1051+2119 & 10:51:48.7890 & +21:19:52.313 & 1.301 & 18.39 & 18.94 & 0.986 & 1.143 & 78.0 & C\tablenotemark{a} & 1474.35\\
 J1054+4541 & 10:54:32.0849 & +45:41:51.889 & 2.523 & 20.03 & 20.19 & 2.038 & 3.784 & 20.9 & X & 50.63 \\
 J1110+3653 & 11:10:05.0375 & +36:53:36.280 & 0.630 & 20.56 & 20.48 & 3.348 & 4.149 & 22.2 & C, X, Ku & 18.62  \\
 J1110+4817 &  11:10:36.3234 & +48:17:52.439 & 0.742\tablenotemark{b} & 20.38 & 20.73 & 3.077 & 6.952 & 17.5 & C\tablenotemark{a} & 541.27\\
 J1137+4825 & 11:37:09.0209 & +48:25:32.616 & 2.261 & 20.15 & 20.14 & 1.951 & 2.750 & 20.3 & X & 51.98\\
 %J1146+3200\tablenotemark{b} & 11:46:36.7544 & +32:00:04.065 & 1.757 & 18.97 & 18.82 & 1.109 & 38.5 & \nodata & 137.35 \\
 J1259+5140 & 12:59:31.1737 & +51:40:56.258 & 1.620 & 19.99 & 20.74 & 5.693 & 6.220 & 25.7 & C\tablenotemark{a} & 275.15 \\
 J1318+3842 & 13:18:36.4854 & +38:42:13.935 & 2.342 & 19.66 & 19.83 & 0.000 & 2.159 & 31.1 & X & 27.99\\
 J1321+3651 & 13:21:42.1532 & +36:51:51.411 & 2.358 & 19.86 & 19.99 & 0.000 & 2.545 & 19.7 & X & 26.31\\
 J1338+1014 & 13:38:34.2190 & +10:14:32.480 & 2.227 & 20.22 & 20.39 & 1.242 & 3.004 & 19.2 & X & 38.52 \\
 J1428+5636 & 14:28:24.7543 & +56:36:11.240 & 2.130 & 19.20 & 19.21 & 1.179 & 1.444 & 62.2 & C\tablenotemark{a} & 109.33 \\
 J1528+1827 & 15:28:00.6115 & +18:27:01.596 & 1.661 & 19.65 & 19.73 & 0.140 & 2.048 & 16.4 & X & 21.32 \\
 J1528$-$0201 & 15:28:17.4573 & $-$02:01:48.597 & 1.366 & 19.88 & 19.77 & 1.276 & 2.005 & 21.6 & X & 22.76 \\
 J1557+1236 & 15:57:59.7015 & +12:36:24.431 & 0.909 & 19.96 & 20.31 & 2.763 & 3.450 & 14.9 & X & 19.81\\
 J1631+4702 & 16:31:40.8857 & +47:02:43.036 & 0.667 & 20.10 & 20.52 & 2.904 & 4.754 & 18.0 & X & 18.76\\
 %J1644+3916 & 16:44:34.4735 & +39:16:04.930 & 1.581 & 19.85 & 19.46 & 1.579 & 28.5 & C\tablenotemark{d}, Ku\tablenotemark{d} & 110.79 \\
 J2339$-$0157 & 23:39:59.0709 & $-$01:57:35.917 & 2.412 & 19.09 & 19.05 & 1.013 & 1.381 & 29.7 & C, X & 36.10 \\
 \enddata
 \tablecomments{Column 1: Target name. Column 2 \& 3: Right Ascension and Declination based on Gaia DR3. Column 4: Spectroscopic redshift from SDSS DR14. Column 5: SDSS $r$-band PSF magnitude. Column 6: Gaia G mean magnitude from DR3. Column 7 \&  8: Gaia {\tt astrometric\_excess\_noise} value from DR3 and DR2. Column 9: Standard deviation of Gaia G-band flux (={\tt phot\_g\_mean\_flux\_error}$\times$({\tt phot\_g\_n\_obs})$^{\frac{1}{2}}$) from DR3. Column 10: VLBA continuum imaging receiver band. Column 11: FIRST integrated flux density at 1.4 GHz. }
 \tablenotetext{a}{Observations from archival VIPS survey \citep{Helmboldt2007}}
% \tablenotetext{b}{Target shows extended radio structure in FIRST.}
 \tablenotetext{b}{The redshift of 6.169 from the SDSS pipeline is not correct.}
 %\tablenotetext{c}{Observations from pilot VIPS survey \citep{Taylor2005}}
\end{deluxetable*}

\section{Observations and Data reduction}\label{sec:observations}

\subsection{New VLBA Observations}\label{sec:new_obs}

%\startlongtable
\begin{deluxetable*}{cccccc}
 \tablecaption{Logs of new VLBA observations for the 18 targets (17 objects without VLBA archival images plus J1044+2959). 
 \label{tab:vlba_obs_info}}
 \tablehead{ \colhead{Name} & \colhead{Band} & \colhead{Observation date } & \colhead{On-source time} & \colhead{Phase cal.} \\ 
 (J2000) & & \colhead{(mm-dd-2020)} & \colhead{(min)} & (J2000)\\
 \colhead{(1)} & \colhead{(2)} & \colhead{(3)} & \colhead{(4)} & \colhead{(5)}
 }
 \startdata
 J0035$-$0911 & X &  02-22, 05-30, 06-01, 06-05, 07-12, 07-27, 08-27 & 168 & J0039$-$0942 \\
 \hline
 \multirow{3}{*}{J0108$-$0400} & X & 02-22, 05-30, 06-01, 06-05, 07-12, 07-27, 08-27 & 168 & \multirow{3}{*}{J0106$-$03} \\
  & C & 07-13, 07-26 & 64\\
  & Ku & 07-04, 07-30 & 144 \\
  \hline
  \multirow{2}{*}{J0238+0123} & X & 02-22, 05-30, 06-01, 06-05, 07-12, 07-27, 08-27 & 168 & \multirow{2}{*}{J0239+04} \\
  & C & 07-13, 07-26 & 80\\
  \hline
  \multirow{3}{*}{J0749+2255} & X & 04-18, 04-24, 05-12, 08-29, 09-11 & 41 & \multirow{3}{*}{J0748+24} \\
  & C & 08-27 & 24\\
  & Ku & 07-03, 08-28 & 120 \\  
    \hline
  J0840+4439 & X & 04-18, 04-24, 05-12, 08-29, 09-11 & 122 &  self \\
  \hline
  \multirow{2}{*}{J1044+2959} & X & 04-18, 04-24, 05-12, 08-29, 09-11 & 125 &  \multirow{2}{*}{self} \\
   & Ku & 07-03, 08-28 & 20 \\ 
  \hline
   J1046+4827 & X & 04-18, 04-24, 05-12, 08-29, 09-11 & 22 & J1027+48 \\ 
  \hline
  J1054+4541 & X & 04-18, 04-24, 05-12, 08-29, 09-11 & 22 & J1058+43 \\
  \hline
  \multirow{3}{*}{1110+3653}  & X & 04-18, 04-24, 05-12, 08-29, 09-11 & 46 & \multirow{3}{*}{J1104+38} \\
  & C & 08-27 & 24 \\ 
  & Ku & 07-03, 08-28 & 120 \\ 
  \hline
  J1137+4825 & X & 04-18, 04-24, 05-12, 08-29, 09-11 & 22 & J1138+47 \\
  \hline
  J1318+3842 & X & 03-03, 03-29, 04-11 & 36 & J1322+39 \\
  \hline
  J1321+3651 & X & 03-03, 03-29, 04-11 & 36 & J1324+36 \\
  \hline
  J1338+1014 & X & 03-03, 03-29, 04-11 & 36 & J1342+12 \\
  \hline
  J1528+1827 & X & 03-03, 03-29, 04-11 & 36 & J1535+19 \\
    \hline
  J1528$-$0201 & X & 03-03, 03-29, 04-11 & 36 & J1533$-$04 \\
    \hline
  J1557+1236 & X & 03-03, 03-29, 04-11 & 36 & J1555+11 \\
  \hline
  J1631+4702 & X & 03-03, 03-29, 04-11 & 36 & J1637+47 \\
  \hline
  \multirow{2}{*}{J2339$-$0157} & X &  02-22, 05-30, 06-01, 06-05, 07-12, 07-27, 08-27 & 168 & \multirow{2}{*}{J2337$-$02}\\
  & C & 07-13, 07-26 & 80\\
 \enddata
 \tablecomments{Column 1: Target name. Column 2: Receiver band. Column 3: Observation date. Column 4: Total on-source time. Column 5: Phase-reference calibrator. The experimental setup for X-band observations includes single polarization, 2 s integration times, and a total bandwidth of 512 MHz (16 spectral channels with a bandwidth of 32 MHz). All targets were observed in the phase-referencing mode, with nearby bright quasars used as the phase-reference calibrators, except for J0840+4439 and J1044+2959, which were self-calibrated.}
\end{deluxetable*}

Among the 23 targets, 18 (17 objects without VLBA archival images plus J1044+2959) were observed with VLBA in X-band (8.37~GHz) between February 2020 and September 2020 (Program ID: BL276; PI: X.~Liu). Additional observations were conducted in C-band (4.87 GHz) and/or Ku-band (15.2 GHz) for targets showing extended structures and/or offsets from the Gaia positions in the X-band images. The experimental setup for X-band observations included single polarization, 2 s integration times, and a total bandwidth of 512 MHz (16 spectral channels with a bandwidth of 32 MHz). The exposure time was chosen based on 1.5 Ghz FIRST flux density assuming a conservative spectral index of -1 for synchrotron radiation, a factor of 0.3 reduce in flux going from arcsecond to VLBA scales, a flux ratio of 1:5 between potentially resolved binary components, and at least a 6-sigma detection for each component. At least 8 stations were used during all observations. The antenna at FD (Fort Davis, TX) was used as the reference antenna. All targets were observed in the phase-referencing mode, with nearby bright quasars used as the phase-reference calibrators, except for J0840+4439 and J1044+2959, which were self-calibrated. A single bright quasar from the following list was used as the amplitude calibrator for each observing session: 3C454.3, 0235+164, 4C39.25, and 3C345. The observation details are summarized in \autoref{tab:vlba_obs_info}.

%We observed in the X, C, and/or Ku bands to optimize both sensitivity and resolution. We request dual-band imaging given the number of potential contaminants. Even a two-point spectrum can distinguish lenses (not a lens if the spectra of two compact sources differ) and pc-scale jets (expected to have steep spectra) from bona-fide dual AGN (SMBHs/cores will have flat spectra). The proposed dual-band imaging for the binary AGN candidate from our preliminary study will provide a useful spectral index at the VLBA resolution and will serve as a double check of the astrometry and image fidelity of the archival VLBA observations.  Our ultimate goal is to observe all the ``radio-bright'' objects in our sample selected with Gaia astrometric excess noise, but here we propose to observe the brightest sources as a pilot study to demonstrate the feasibility and efficiency of our Gaia-based astrometry selection technique. Another advantage of observing these brightest targets is that they would not require in-beam calibration, which greatly simplifies the observing strategy. We request phase referencing to get positions. We need two epochs (at different hour angles) for image fidelity. Exposure time estimates are detailed in the Technical Justification.

\subsection{Archival VLBA Observations}

%Among the 23 targets with high Gaia astrometric excess noise, 6 candidates (including J1044+2959) have archival C-band images in the VLBA archive, mainly from the VLBA Imaging and Polarization Survey \citep[VIPS;][]{Helmboldt2007}. 

Of the 23 targets, 6 (including J1044+2959) have archival VLBA C-band images from the VLBA Imaging and Polarization Survey \citep[VIPS;][]{Helmboldt2007} and have been re-calibrated with improved positions from \citet{Petrov2011}. 
The VIPS observations were conducted in 2006 at a center frequency of 5~GHz, with a typical 1$\sigma$ rms noise level of~0.2~mJy~beam${}^{-1}$. The improved absolute astrometry from \citet{Petrov2011} has a position uncertainty of 0.5~mas. Of the six targets, only J1044+2959 shows a significant VLBA-Gaia offset and has a compact core structure. Therefore, we propose additional X-band and Ku-band observations for J1044+2959 to its radio spectral indices.

\subsection{Data Reduction and Analysis}

The data reduction and analysis were conducted using the Astronomical Image Processing System \citep[AIPS;][]{Greisen2003}. We followed the standard reduction pipeline task {\sc vlbarun}, which includes corrections to the Earth orientation parameters, ionospheric correction, amplitude correction, and phase correction. Subsequently, we used {\sc dbcon} to combine the visibility data from different observing sessions for the same targets. Then, we created dirty images and used {\sc imagr} to deconvolve (clean) the images. To optimize the sensitivity, we used a natural weighting scheme. The typical minor axes of the synthesized beams were 1.5~mas for C-band, 1~mas for X-band, and~0.5~mas for Ku-band. The pixel scales were 0.3~mas for C-band, 0.2~mas for X-band, and 0.1~mas for Ku-band, which were chosen to have at least five pixels in the synthesized beam. Additionally, we created wide-field images to search for any radio cores up to~0\farcs6, which is the typical resolving power of Gaia.

%Then, we imaged and cleaned dirty images with {\sc imagr}. We used a natural weighting scheme for source identification as it gives higher sensitivity. The pixel scale are 0.3 mas for C-band, 0.2 mas for X-band, and 0.1 mas for Ku-band. We also made wide-field images to search for any cores up to 0.6 arcsec, which is the typical resolving power of Gaia.

%\blue{Please describe VLBA data reduction and data analysis.}

%\blue{Please fill in all the tables and adjust the measurement columns as needed.}

%\blue{Please add in figures showing all the new VLBA detection as well as relevant archival VLBA observations (e.g., for 1044).}

%\blue{(Once all the measurements and figures are added in, we can start filling in the results and discussion sections.)}

%%%%%%%%%%%%%%%%%%%%%%%%%%%%%%%%%%%%%
%% VLBA X-band detected targets:
%%%%%%%%%%%%%%%%%%%%%%%%%%%%%%%%%%%%%

\begin{deluxetable*}{lccccccccccc}
  \tablecaption{Source and image properties of the 23 targets having the new and archival VLBA observations \label{tab:fitting}}
  %\tablewidth{0pc}
  \tablewidth{\textwidth}
  \tablehead{%
\colhead{Name} & 
\colhead{Band} & 
\colhead{R.A.$_{\rm VLBA}$} & \colhead{Decl.$_{\rm VLBA}$} & 
\colhead{$S_p$} & 
\colhead{$S_T$} &
\colhead{Noise Level} & 
%\colhead{} & 
%\colhead{$S_{{\rm FIRST}}$} &
%\colhead{$\sigma_{{\rm FIRST}}$} &
\colhead{$\theta_{\rm maj}$} &
\colhead{$\theta_{\rm min}$} & 
\colhead{P.A.}
\\
\colhead{} &
\colhead{} &
\colhead{(J2000)} & \colhead{(J2000)}  &
\colhead{(mJy beam$^{-1}$)} & 
\colhead{(mJy)} & 
\colhead{(mJy beam$^{-1}$)} &
%\colhead{Redshift} & 
%\colhead{(mJy)} &
%\colhead{(\mjybm)} &
\colhead{(mas)} & 
\colhead{(mas)} & 
%\colhead{(\mjybm)} & 
\colhead{(degree)} & 
%\colhead{(\mjybm)} & 
%\colhead{($10^{23}$~$\frac{{\rm W}}{{\rm Hz}}$)} 
\\
\colhead{(1)} &
\colhead{(2)} & 
\colhead{(3)} & 
\colhead{(4)} & 
\colhead{(5)} & 
\colhead{(6)} & 
\colhead{(7)}  & 
\colhead{(8)}  & 
\colhead{(9)}  & 
\colhead{(10)}  & 
        }
\startdata
% flux densities determined from CLEAN components
% nothing else within 2'' x 2''
J0035$-$0911 & X  & 00:35:53.0323 & -09:11:50.152 & 0.78 & 1.12 & 0.05 & 3.6 & 1.1 & -12.7\\
\hline
\multirow{3}{*}{J0108$-$0400} & C & 01:08:28.1866 & -04:00:44.998 & 8.00 & 12.81 & 0.19 & 2.6 & 1.0 & 6.8\\
& X & 01:08:28.1866 & -04:00:44.998 & 5.91 & 7.45 & 0.17 & 2.9 & 1.0 & -8.0\\
& Ku & 01:08:28.1866 & -04:00:44.998 & 1.25 & 3.58 & 0.13 & 1.5 & 0.5 & -8.4\\
\hline
\multirow{2}{*}{J0238+0123} & C & 02:38:22.4506 & +01:23:53.212 & 4.56 & 6.37 & 0.11 & 4.0 & 1.6 & 9.6\\
& X & 02:38:22.4507 & +01:23:53.213 & 2.67 & 4.14 & 0.16 & 2.9 & 1.1 & -11.2\\
\hline
J0740+2852 & C\tablenotemark{a} & 07:40:33.5438 & +28:52:47.246 & \nodata & 120.0 & \nodata & 3.0 & 2.2 & 0.0\\
\hline
\multirow{3}{*}{J0749+2255} & C & 07:49:22.9665 & +22:55:11.765 & 23.73 & 34.60 & 0.49 & 3.7 & 1.4 & -2.3\\
& X & 07:49:22.9665 & +22:55:11.766 & 19.46 & 39.50 & 0.55 & 2.5 & 1.0 & -12.9\\
& Ku & 07:49:22.9665 & +22:55:11.767 & 3.39 & 7.79 & 0.19 & 1.1 & 0.5 & -11.8\\
\hline
J0840+4439 & X & 08:40:51.2889 & +44:39:59.291 & 26.55 & 34.66 & 0.61 & 3.7 & 2.5 & -3.6\\
\hline
\multirow{3}{*}{J1044+2959} & C\tablenotemark{a} & 10:44:06.3428 & +29:59:01.004 & \nodata & 145.0 & \nodata & 3.0 & 1.8 & 0.0\\
& X & 10:44:06.3428 & +29:59:01.004 & 127.44 & 135.07 & 0.67 & 2.6 & 1.1 & -0.9\\
& Ku & 10:44:06.3428 & +29:59:01.004 & 2.23 & 2.18 & 0.32 & 3.4 & 0.3 & -8.7\\
\hline
J1046+4827 & X & 10:46:57.1588 & +48:27:23.794 & 1.94 & 3.92 & 0.08 & 2.3 & 1.1 & 14.0\\
\hline
J1051+2119 & C & 10:51:48.7890 & +21:19:52.313 & \nodata & 980.0 & \nodata & 3.0 & 1.8 & 0.0\\
\hline
J1054+4541 & X & 10:54:32.0849 & +45:41:51.890 & 2.16 & 2.70 & 0.11 & 2.3 & 1.1 & 8.7\\
\hline
\multirow{3}{*}{J1110+3653} & C &  11:10:05.0377 & +36:53:36.278 & 2.49 & 2.44 & 0.09 & 4.2 & 1.6 & 28.3\\
& X & 11:10:05.0377 & +36:53:36.278 & 8.62 & 10.60 & 0.15 & 2.4 & 1.1 & 5.9\\
& Ku & 11:10:05.0377 & +36:53:36.278 & 1.36 & 1.75 & 0.09 & 1.1 & 0.5 & 7.4\\
\hline
J1110+4817 & C\tablenotemark{a} & 11:10:36.3241 & +48:17:52.449 & \nodata & 181.0 & \nodata & 3.0 & 1.8 & 0.0\\
\hline
J1137+4825 & X & 11:37:09.0209 & +48:25:32.616 & 42.61 & 54.73 & 0.92 & 2.4 & 1.2 & 13.0\\
\hline
J1259+5140 & C\tablenotemark{a} & 12:59:31.1739 & +51:40:56.260 & \nodata & 264.0 & \nodata & 3.0 & 1.8 & 0.0\\
\hline
J1318+3842 & X & \nodata & \nodata & \nodata & $<$0.36 & 0.07 & 2.7 & 1.8 & 0.2\\
\hline
J1321+3651 & X & 13:21:42.1532 & +36:51:51.411 & 3.04 & 4.84 & 0.35 & 2.9 & 1.9 & -3.2\\
\hline
J1338+1014 & X & 13:38:34.2190 & +10:14:32.480 & 1.89 & 1.73 & 0.20 & 4.1 & 2.4 & -4.3\\
\hline
J1428+5636 & C\tablenotemark{a} & 14:28:24.7543 & +56:36:11.240 & \nodata & 139.0 & \nodata & 3.0 & 2.2 & 0.0\\
\hline
J1528+1827 & X & \nodata & \nodata & \nodata & $<$0.80 & 0.16 & 3.7 & 2.4 & 0.0\\
\hline
J1528$-$0201 & X & 15:28:17.4572 & -02:01:48.598 & 0.32 & 0.83 & 0.06 & 3.7 & 1.6 & -4.5\\
\hline
J1557+1236  & X & 15:57:59.7009 & +12:36:24.370 & 0.39 & 1.56 & 0.06 & 3.0 & 1.4 & -4.4\\
\hline
J1631+4702 & X & 16:31:40.8857 & +47:02:43.038 & 13.02 & 18.74 & 0.25 & 2.6 & 1.3 & 14.6\\
\hline
\multirow{2}{*}{J2339$-$0157} & C & 23:39:59.0710 & -01:57:35.917 & 4.44 & 5.77 & 0.08 & 4.3 & 1.8 & 12.6\\
& X & 23:39:59.0709 & -01:57:35.917 & 1.48 & 2.46 & 0.07 & 2.8 & 1.0 & -9.1\\
\hline
\enddata
% flux densities determined from CLEAN components
\tablenotetext{a}{Measurements from \citet{Petrov2011}.}
\tablecomments{
%Column (1): SDSS designation with J2000 coordinates.
%Column (2): SDSS spectroscopic redshift. 
%Column (3): FIRST integrated flux density at 20 cm; N/A means source is not covered by the FIRST survey.
%Column (4): RMS noise at 20 cm in the FIRST survey map; N/A means source is not covered by the FIRST survey.
Column (1): Target name. Column (2): Receiver band. Column (3) \& (4): Right Ascension and Declination of the VLBA detected source. The position corresponds to the brightest VLBA component in the presence of multiple components. Column (5) \& (6): Peak and total flux density of the detected source. The 5-$\sigma$ upper limit is reported if a source is undetected. Column (7): 1-$\sigma$ noise level. Column (8) \& (9): Major and Minor axis of the synthesized beam. Column (10): Position angle of the synthesized beam. 
}
\end{deluxetable*}
%}

%%%%%%%%%%%%%%%%%%%%%%%%%%%%%%%%%%%%%
%% VLBA X-band undetected targets:
%%%%%%%%%%%%%%%%%%%%%%%%%%%%%%%%%%%%%

%%%
%%%%
%%%%%
%\newgeometry{margin=1cm} % modify this if you need even more space
%\begin{landscape}
%\begin{table}
%\begin{tabular}
% \end{tabular}
%\end{table}
%\end{landscape}
%\restoregeometry
%%%%
%%%
%%

\section{Results}\label{sec:results}

\subsection{VLBA Continuum Images}\label{sec:images}

\begin{figure*}[!h]
    \centering
    \fbox{\includegraphics[width=0.23\textwidth]{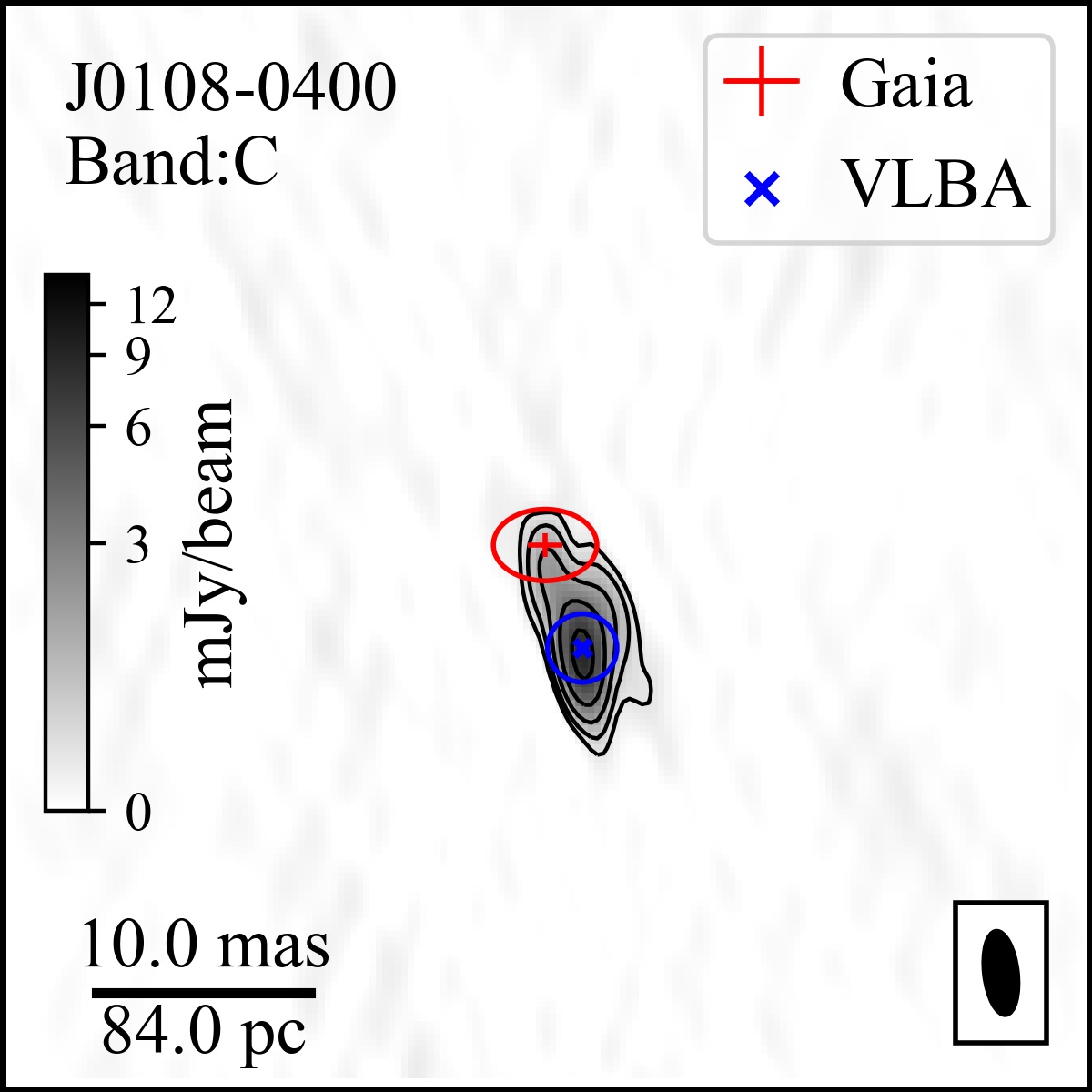}
    \includegraphics[width=0.23\textwidth]{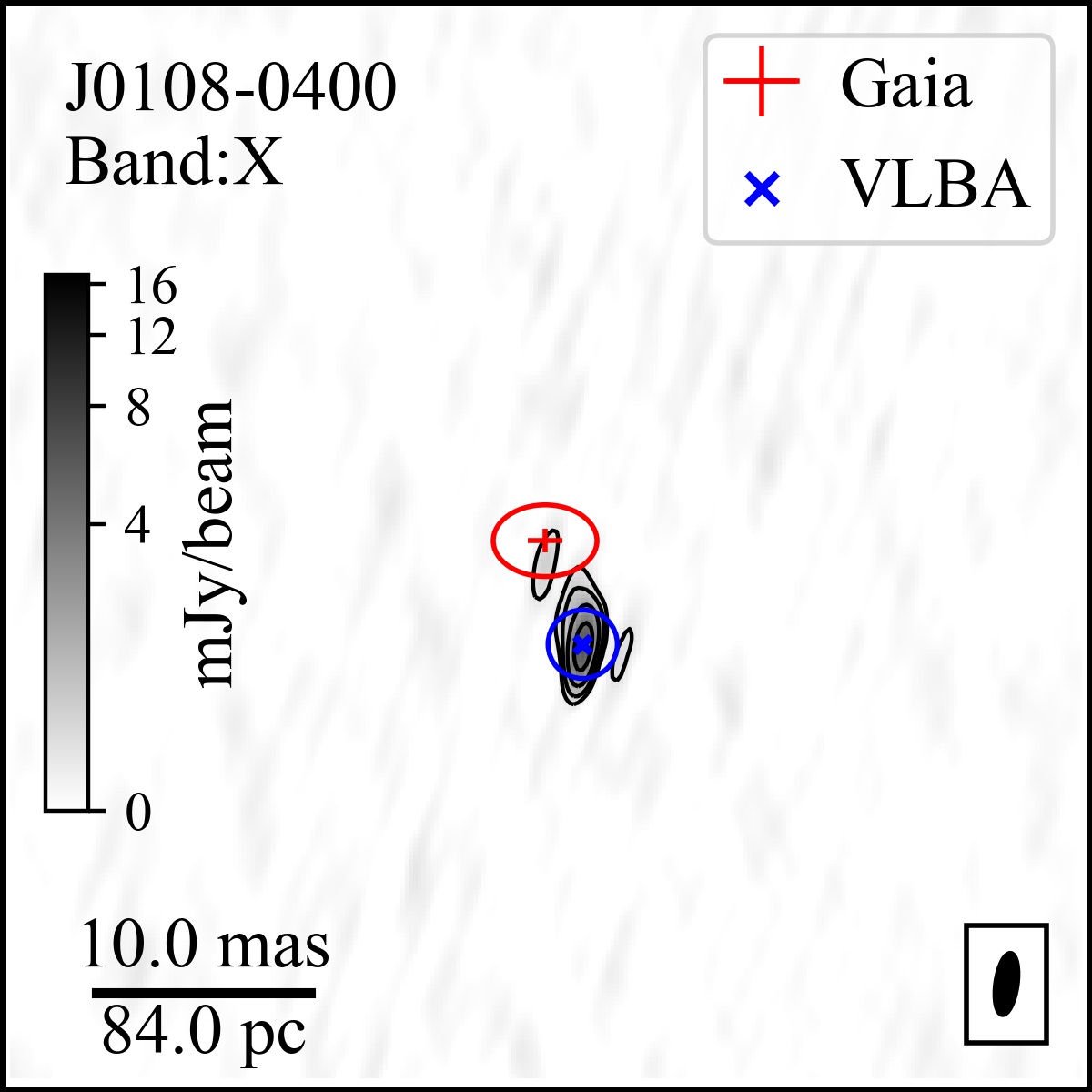}
    \includegraphics[width=0.23\textwidth]{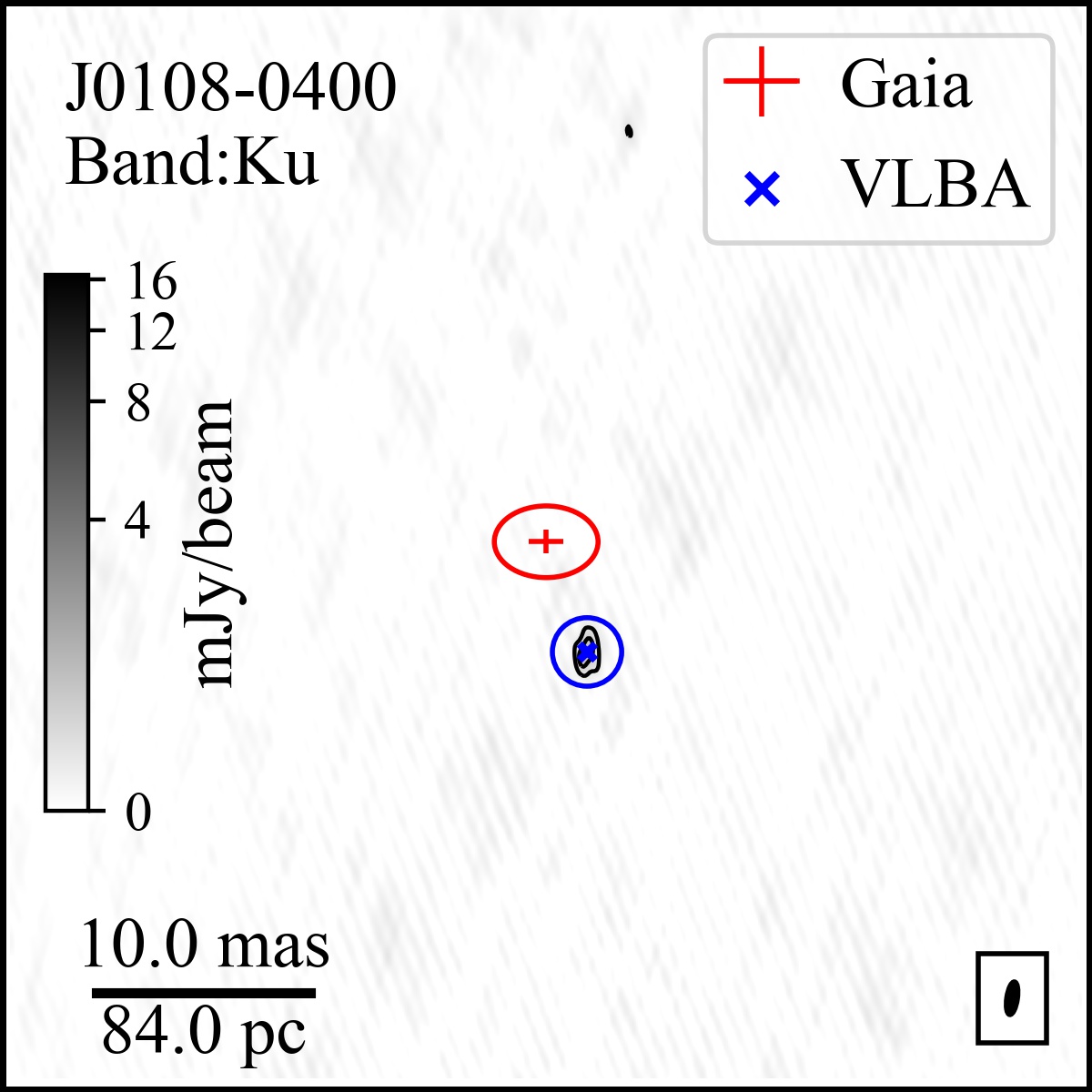}}
    \fbox{\includegraphics[width=0.23\textwidth]{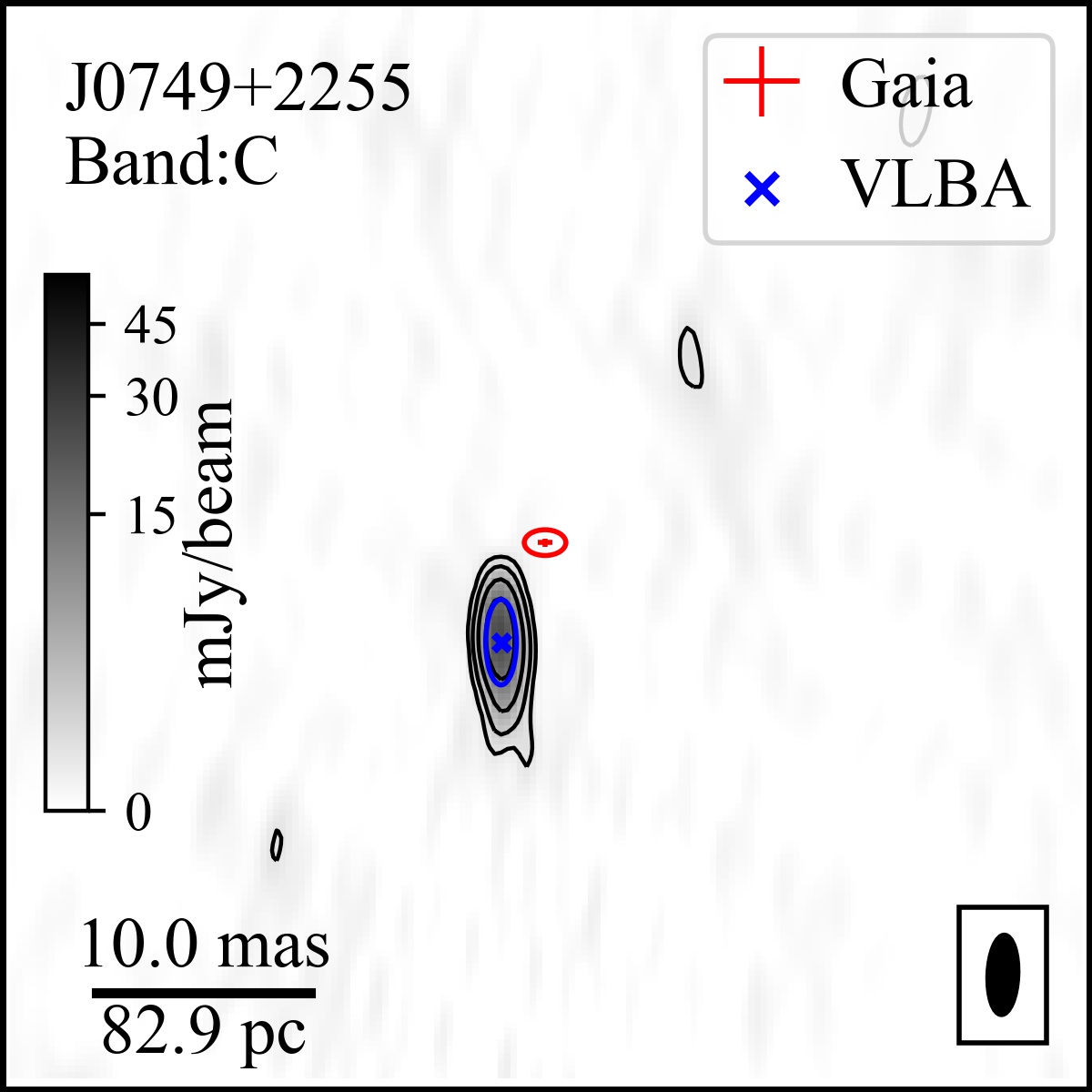}
    \includegraphics[width=0.23\textwidth]{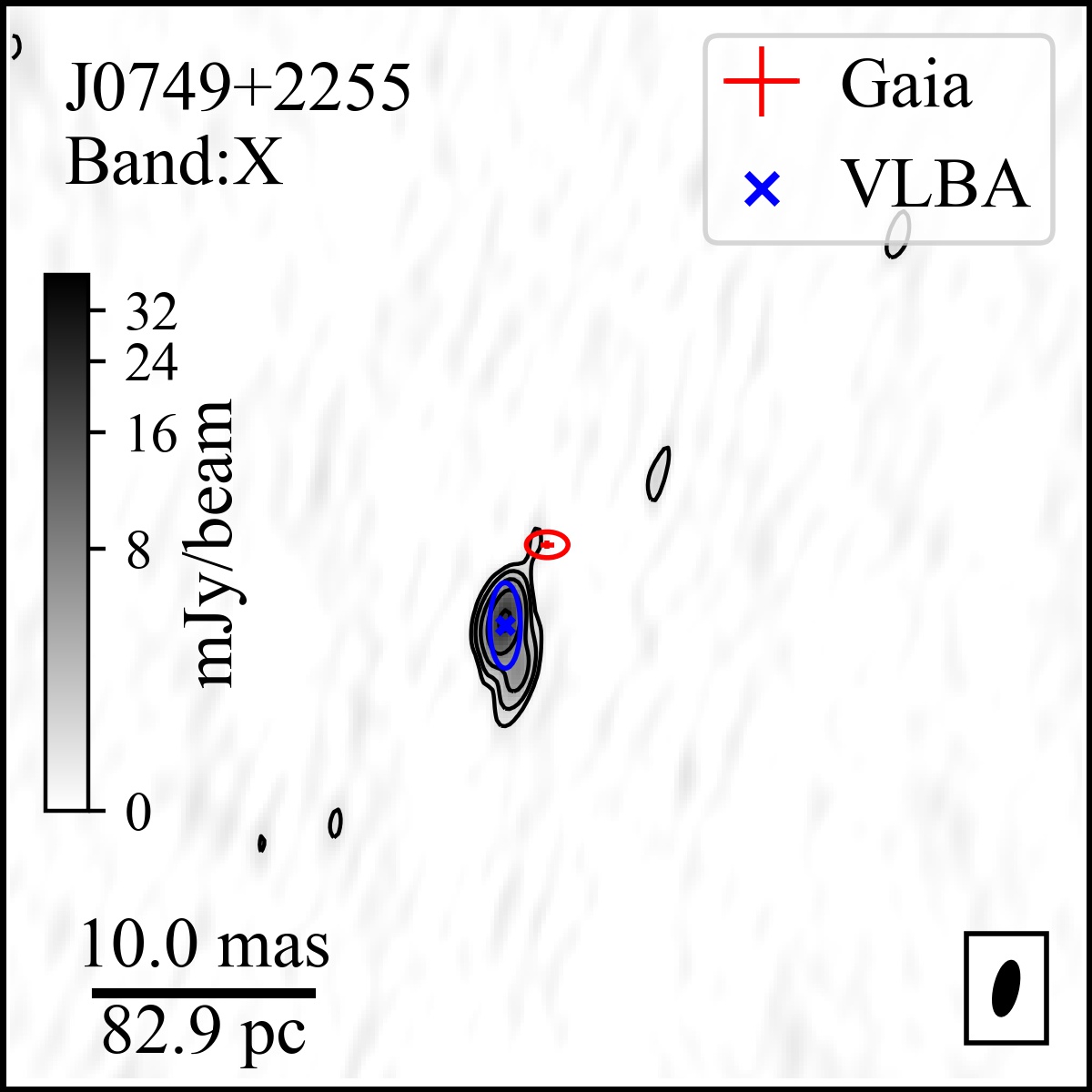}
    \includegraphics[width=0.23\textwidth]{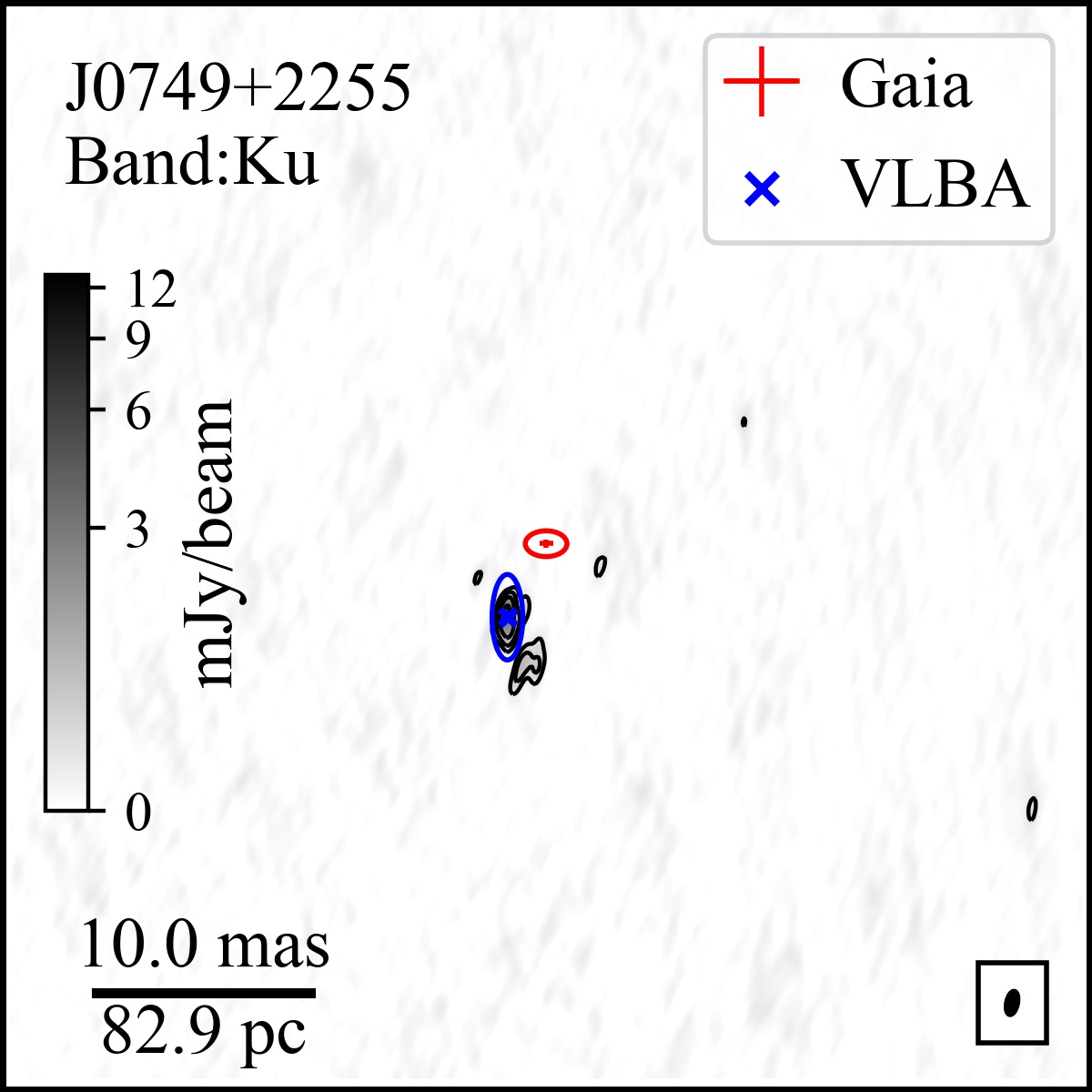}}
    \fbox{\includegraphics[width=0.23\textwidth]{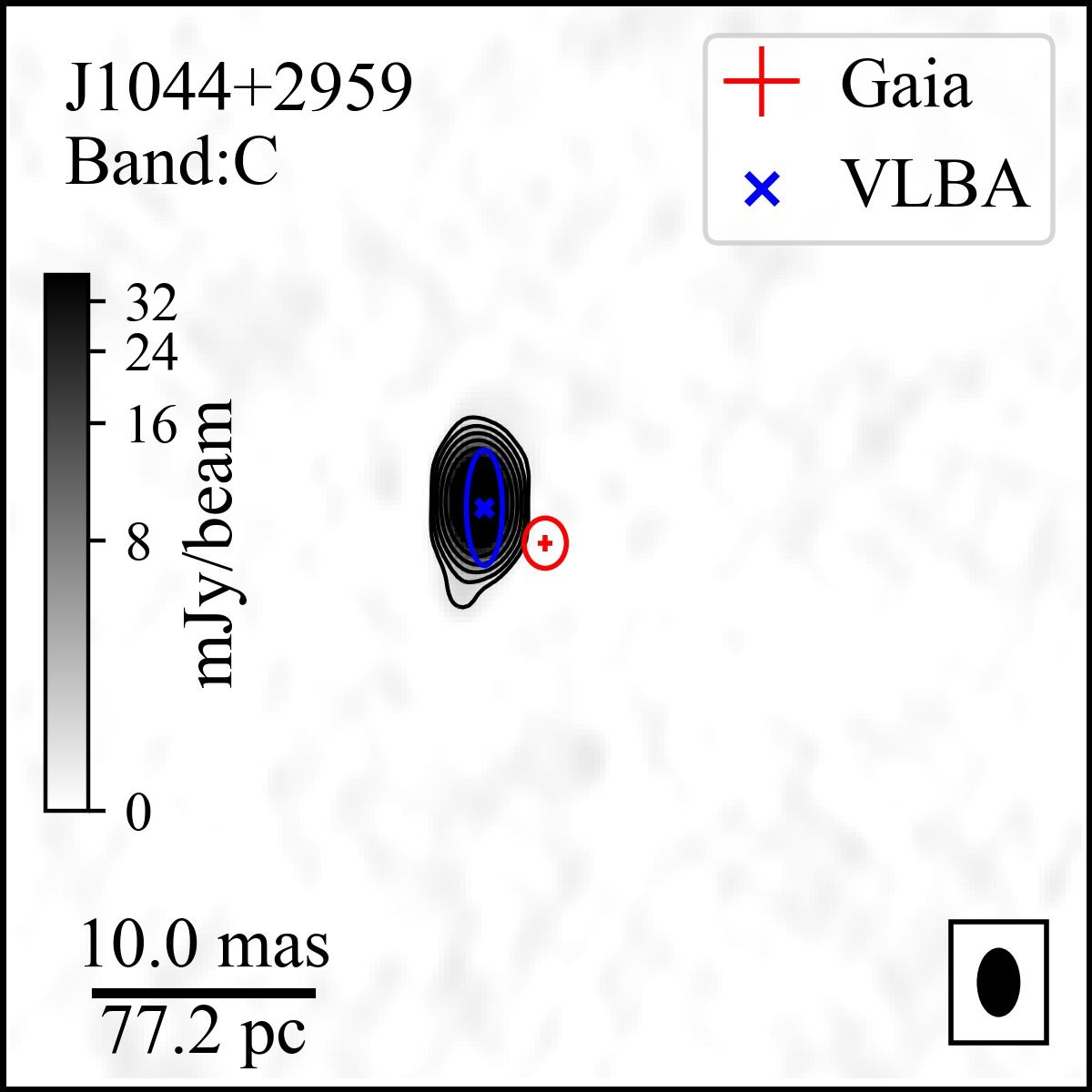}
    \includegraphics[width=0.23\textwidth]{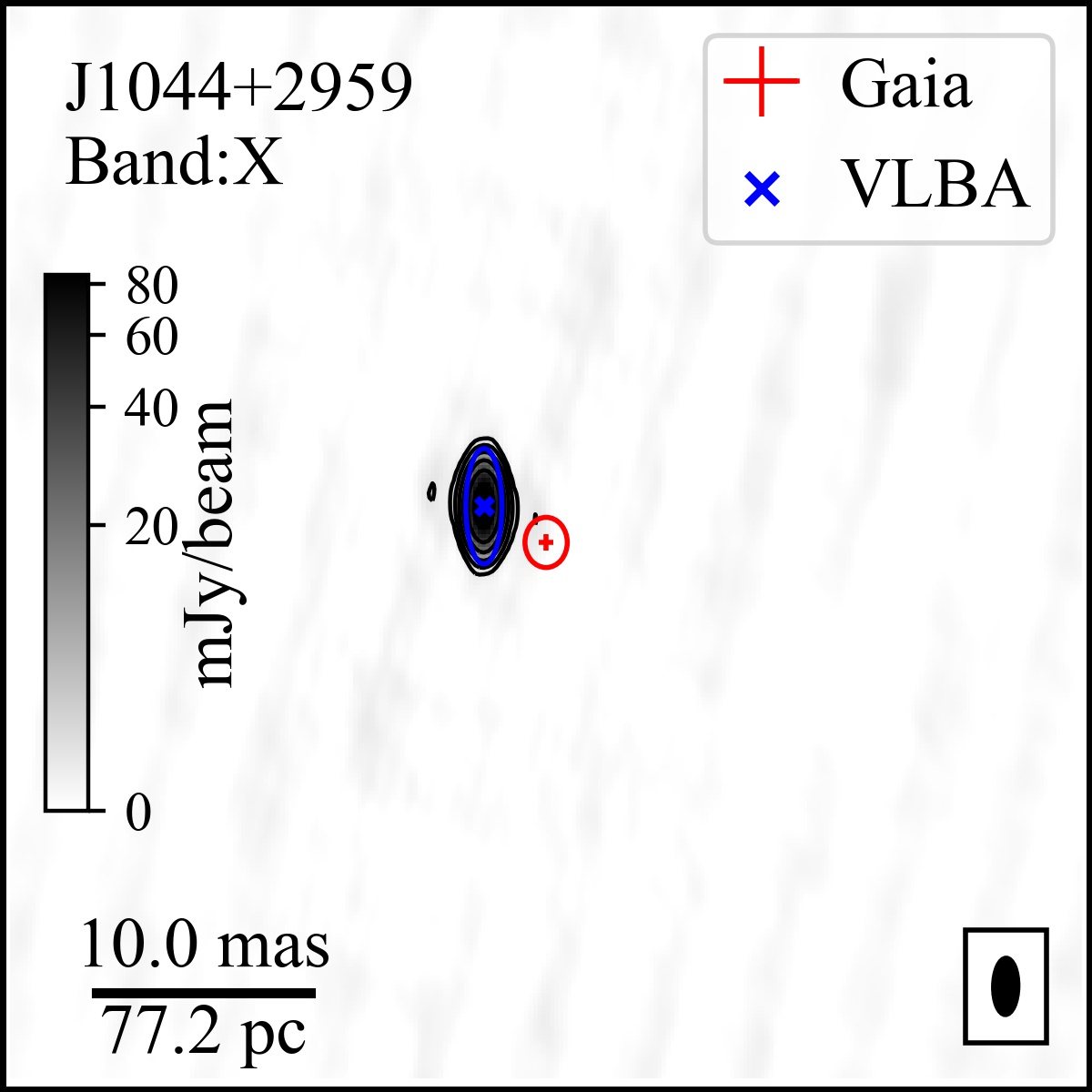}
    \includegraphics[width=0.23\textwidth]{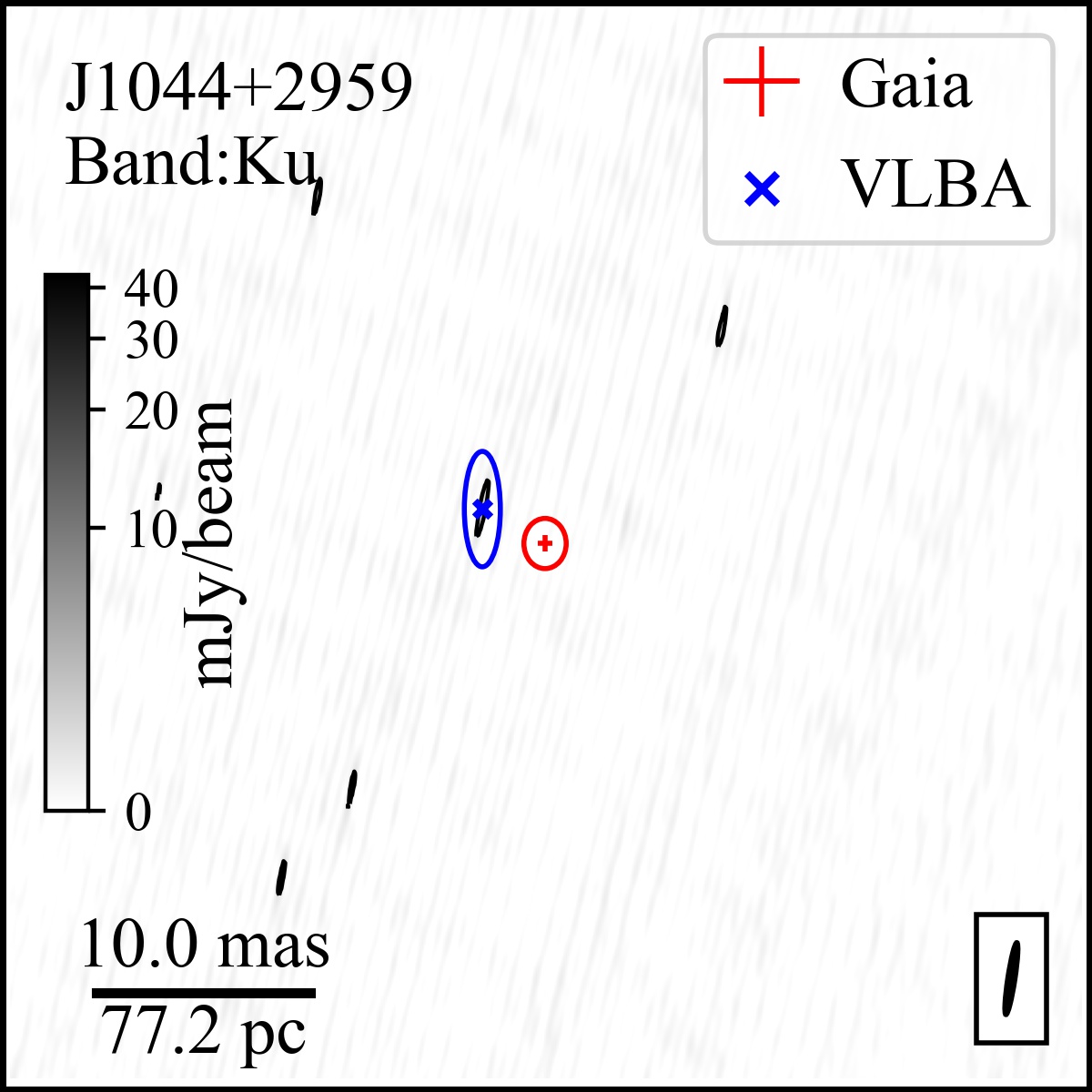}} 
    \\
    \includegraphics[width=0.23\textwidth]{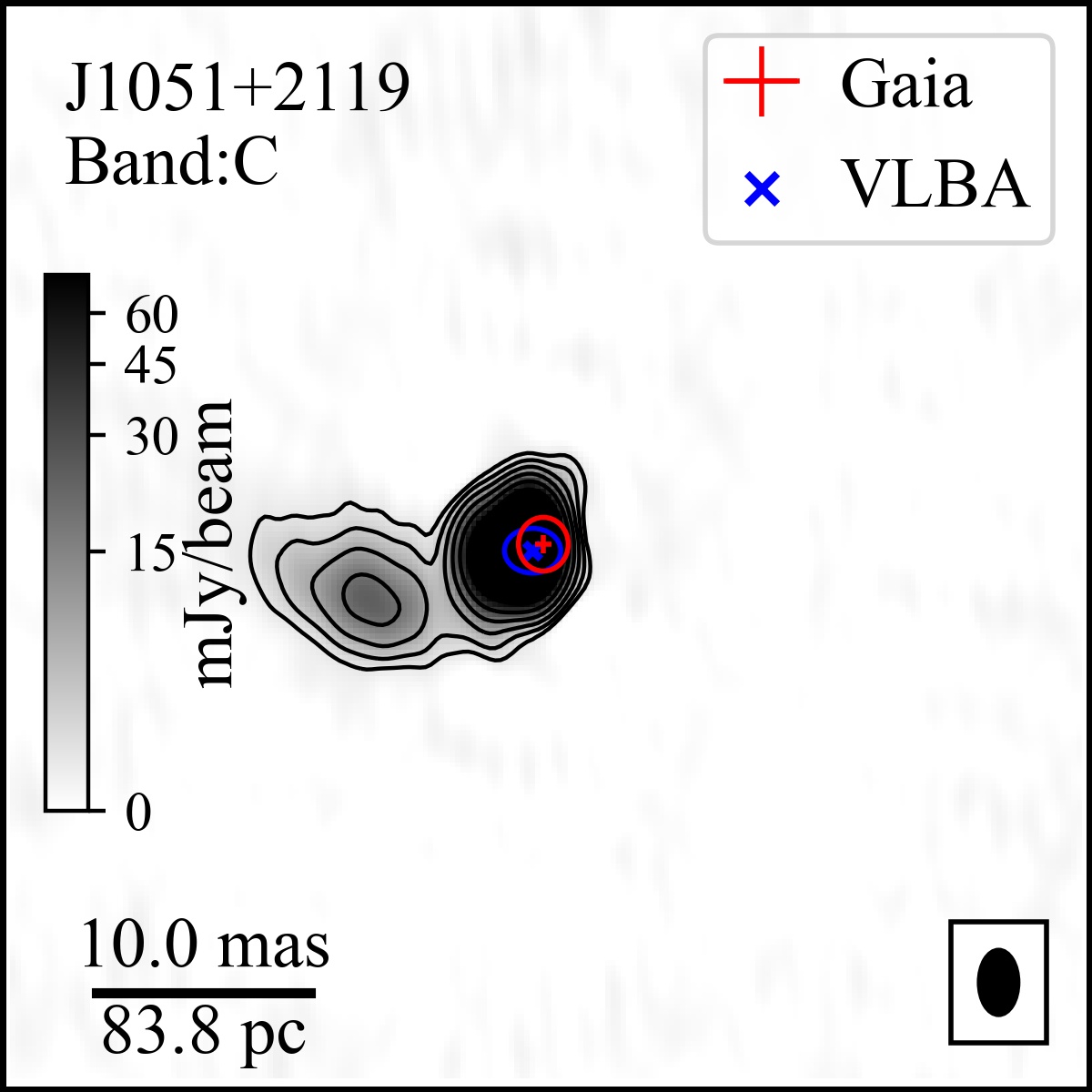}
    
    \caption{
    %\textbf{\color{red} Please add colorbars to show surface brightness (similar to your Nature paper VLA figure). Quote the typical rms noise levels (i know we list them in the table, but it would be good to be reminded about the typical sensitivity reached also directly in the figure caption). Same for Figure 4.} \textbf{\color{blue} YCC:Done}
    VLBA (C-band, X-band, and/or Ku-band) images of the eight (four in this figure and four in Figure \ref{fig:VLBA_images_offset_2}) targets that exhibit significant offsets between VLBA and Gaia positions or/and have multiple radio components. The images are oriented with north up and east to the left. The Gaia DR2 positions are indicated by red crosses, with their lengths representing 1$\sigma$ rms errors, and the red ellipses show the 3$\sigma$ rms position error. The VLBA positions are marked by blue crosses, and the blue ellipses indicate the 3$\sigma$ rms position error. The synthesized beam sizes are shown in the bottom-right corners of the images. The solid contours correspond to~5, 10, 20, 40, 80, 160, and 320 times the 1$\sigma$ rms level. The typical 1$\sigma$ rms noise level is approximately 0.1~mJy~beam$^{-1}$, but for bright sources, it can reach up to 1~mJy~beam$^{-1}$.}
    \label{fig:VLBA_images_offset_1}
\end{figure*}

\begin{figure*}
    \centering
    \fbox{\includegraphics[width=0.23\textwidth]{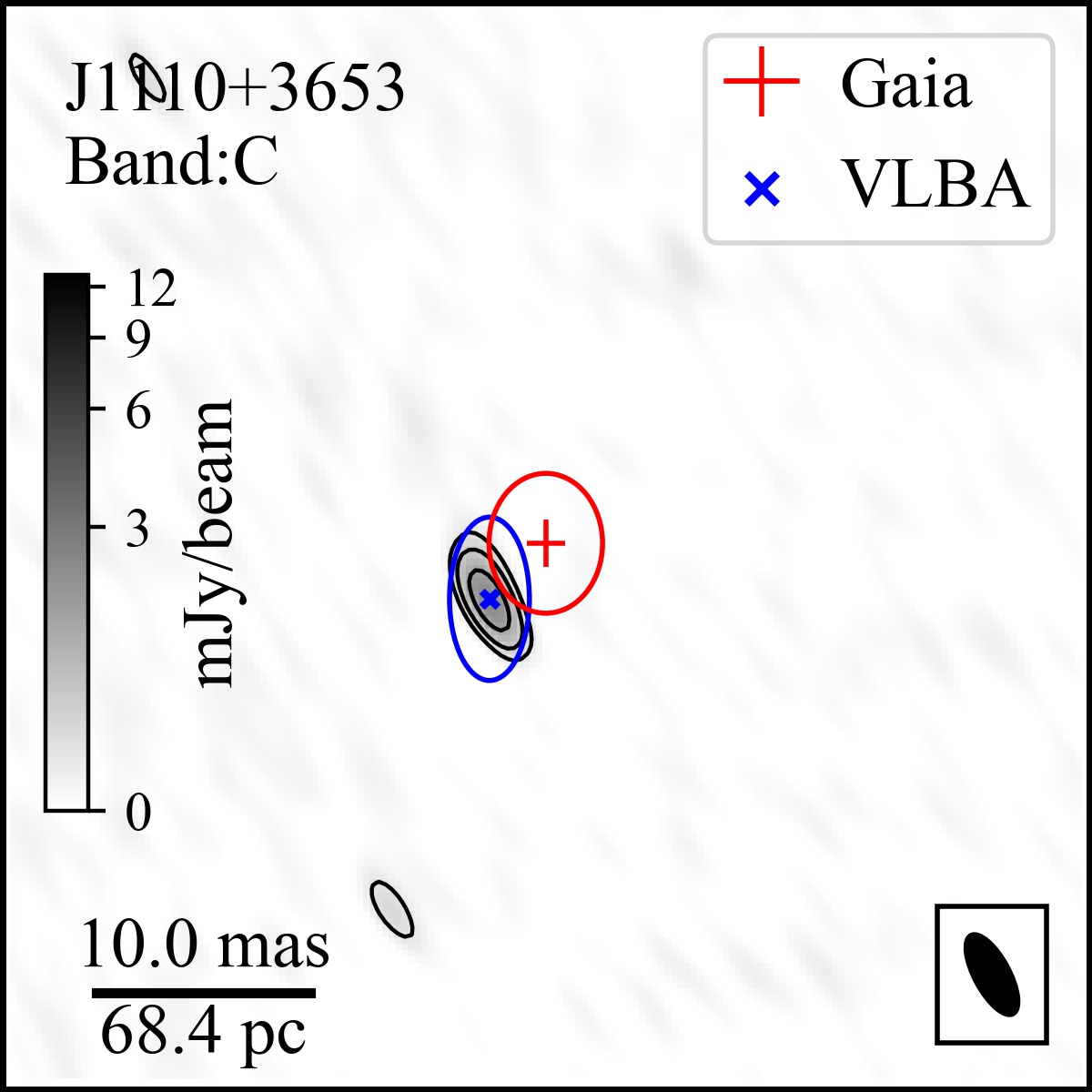}
    \includegraphics[width=0.23\textwidth]{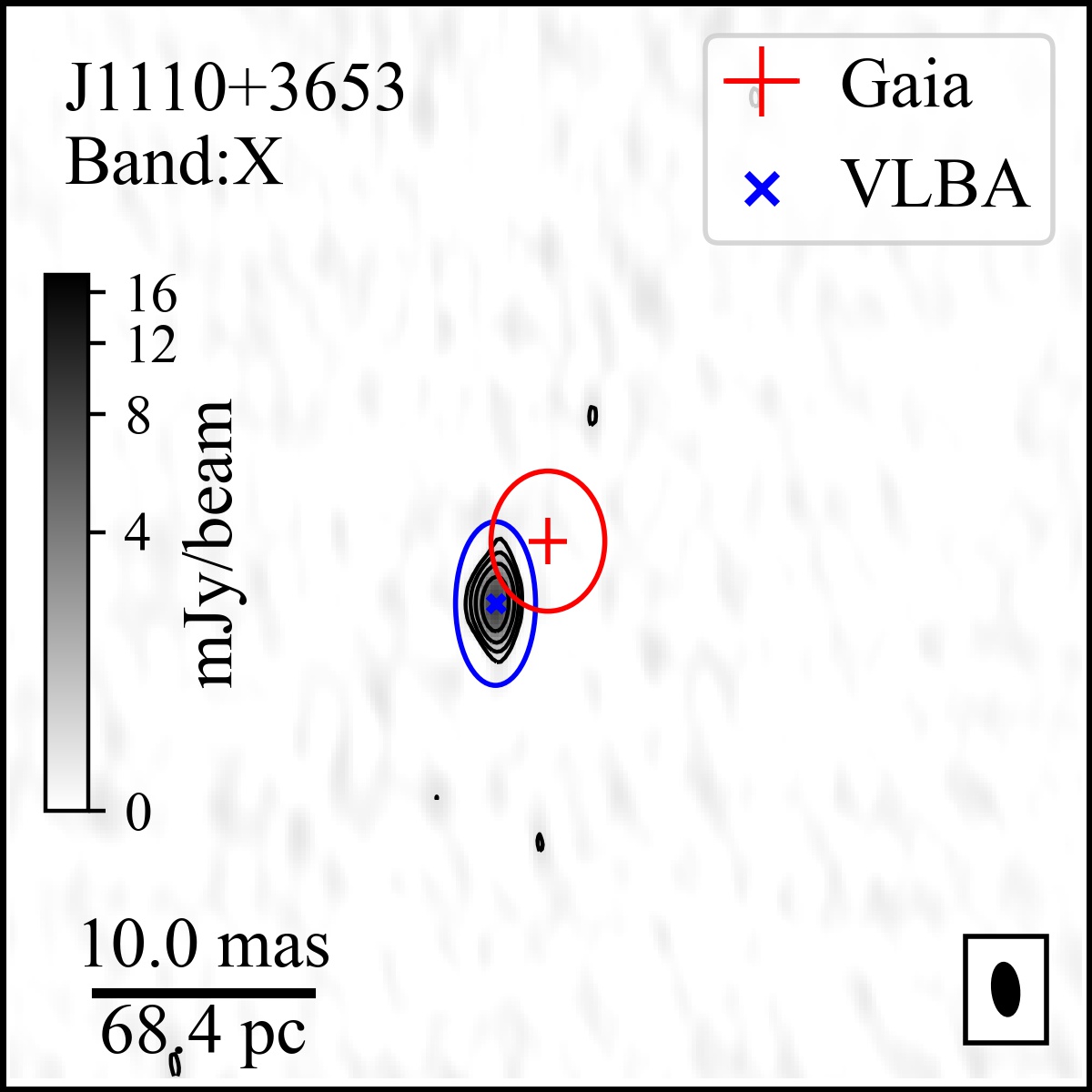} 
    \includegraphics[width=0.23\textwidth]{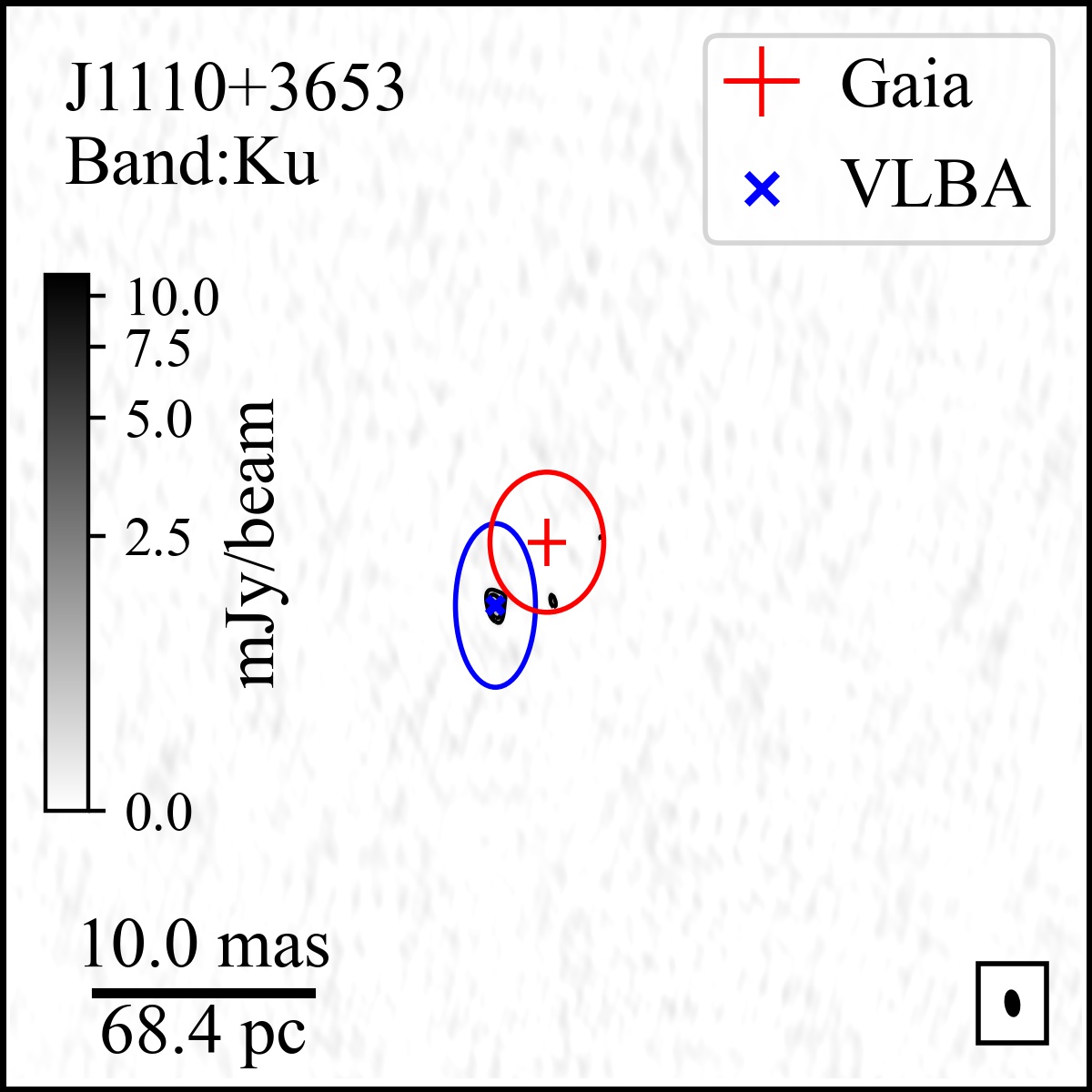}} \\
    \includegraphics[width=0.22\textwidth]{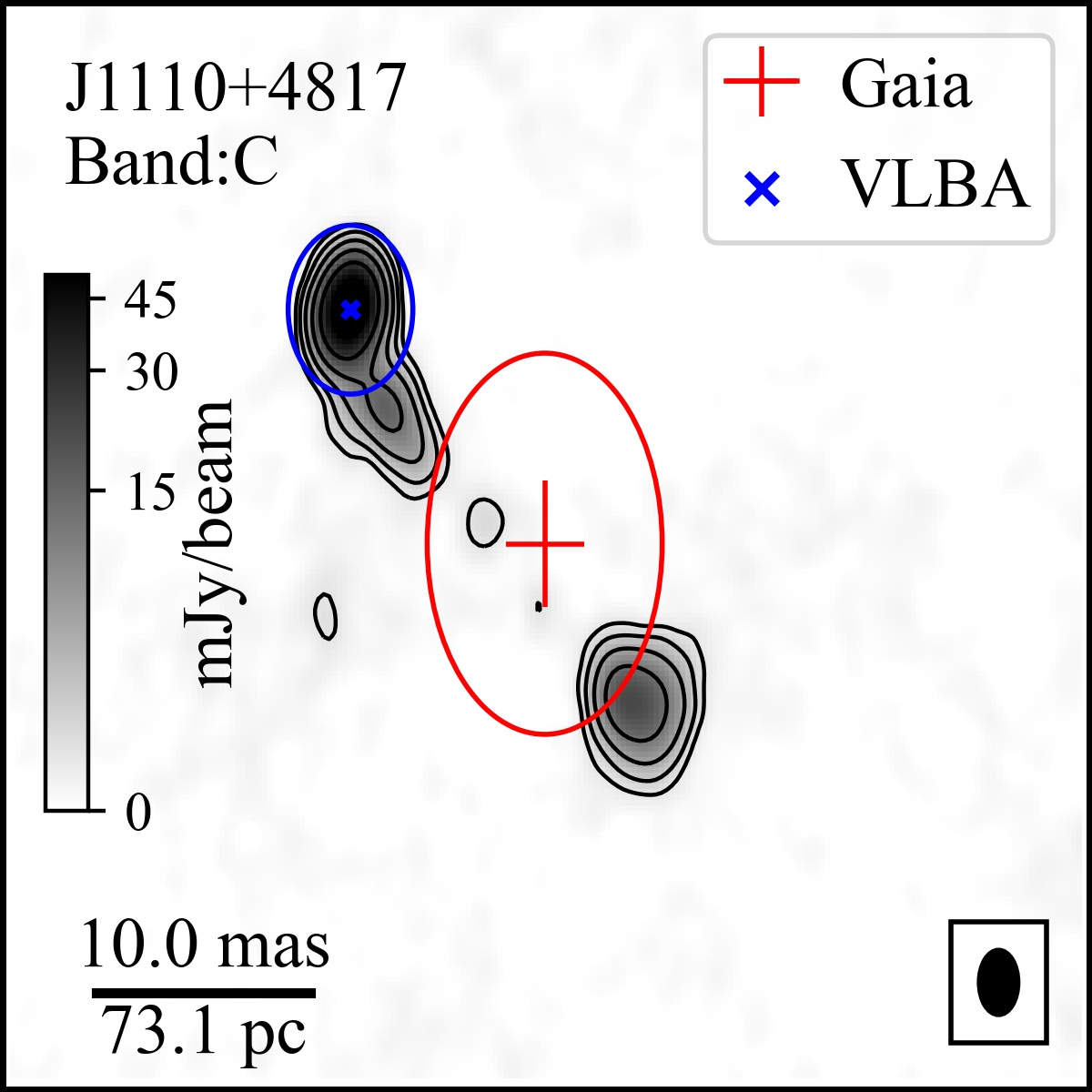}\\
    \includegraphics[width=0.22\textwidth]{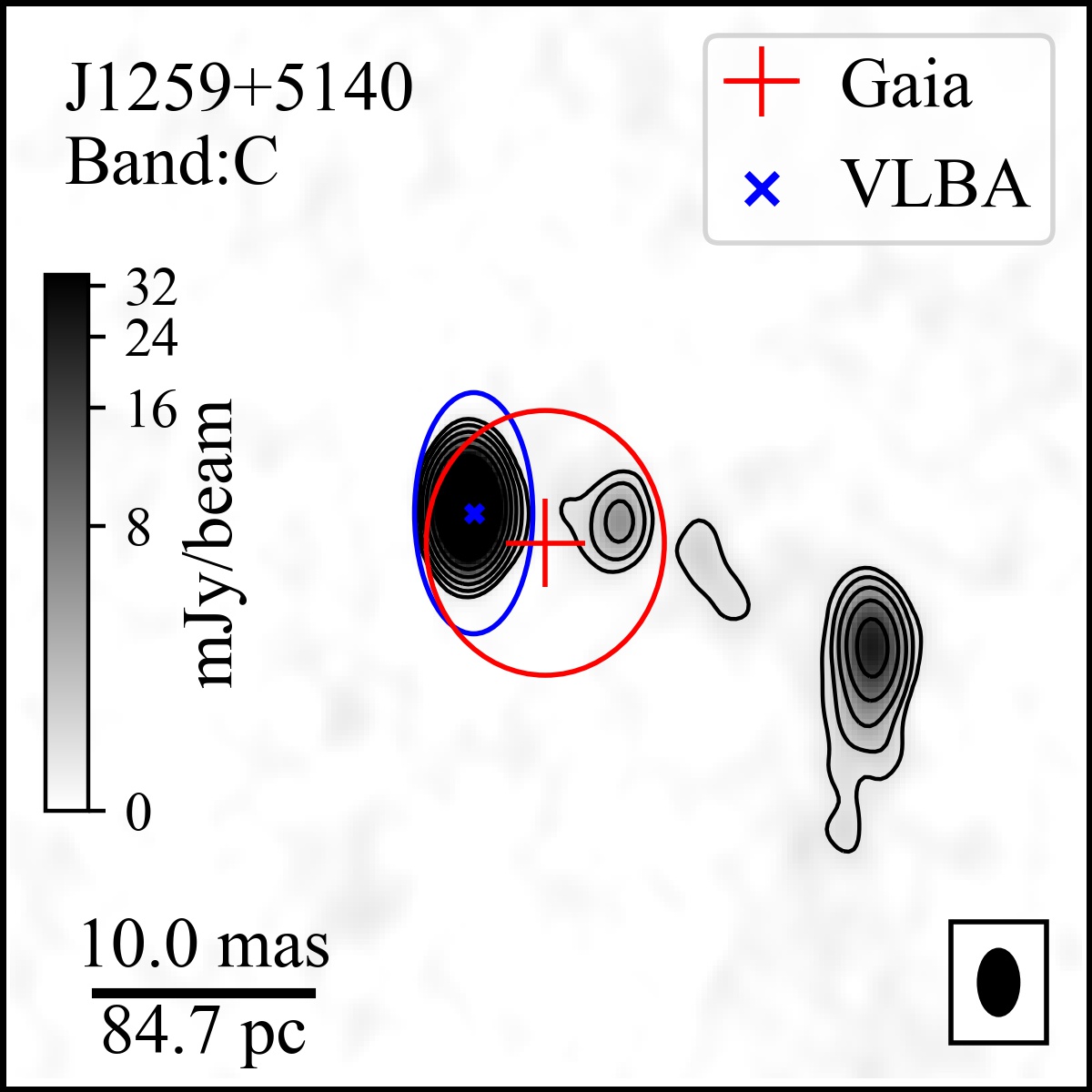}\\
    \includegraphics[width=0.22\textwidth]{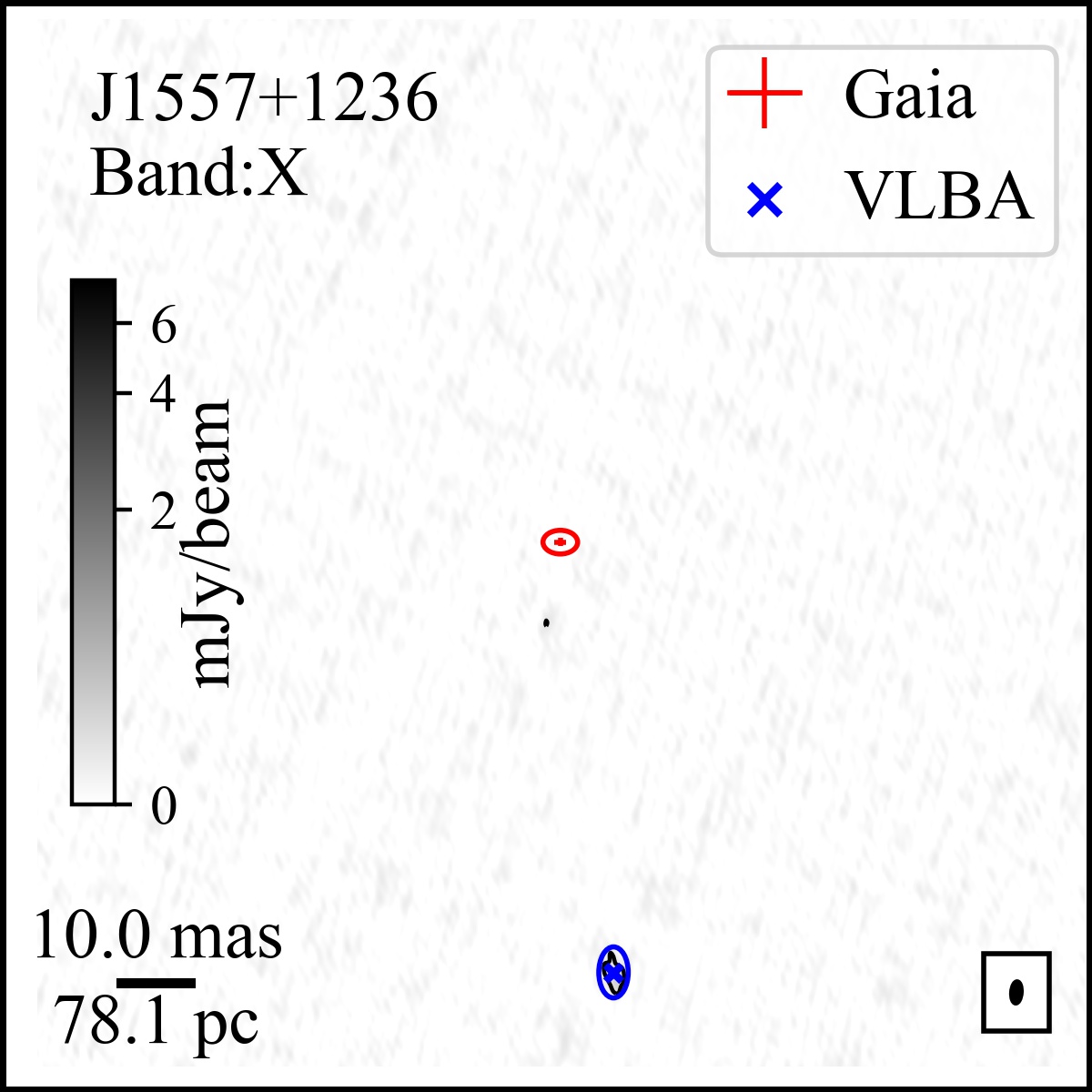}
    \caption{
    %\textbf{\color{red} Please add colorbars to show surface brightness (similar to your Nature paper VLA figure). Quote the typical rms noise levels (i know we list them in the table, but it would be good to be reminded about the typical sensitivity reached also directly in the figure caption). Same for Figure 4.} \textbf{\color{blue} YCC:Done}
    VLBA (C-band, X-band, and/or Ku-band) images of the eight (four in Figure \ref{fig:VLBA_images_offset_1} and four in this figure) targets that exhibit significant offsets between VLBA and Gaia positions or/and have multiple radio components. Notations are the same as those in Figure \ref{fig:VLBA_images_offset_1}.}
    \label{fig:VLBA_images_offset_2}
\end{figure*}

\begin{figure*}[!h]
    \centering

    \includegraphics[width=0.22\textwidth]{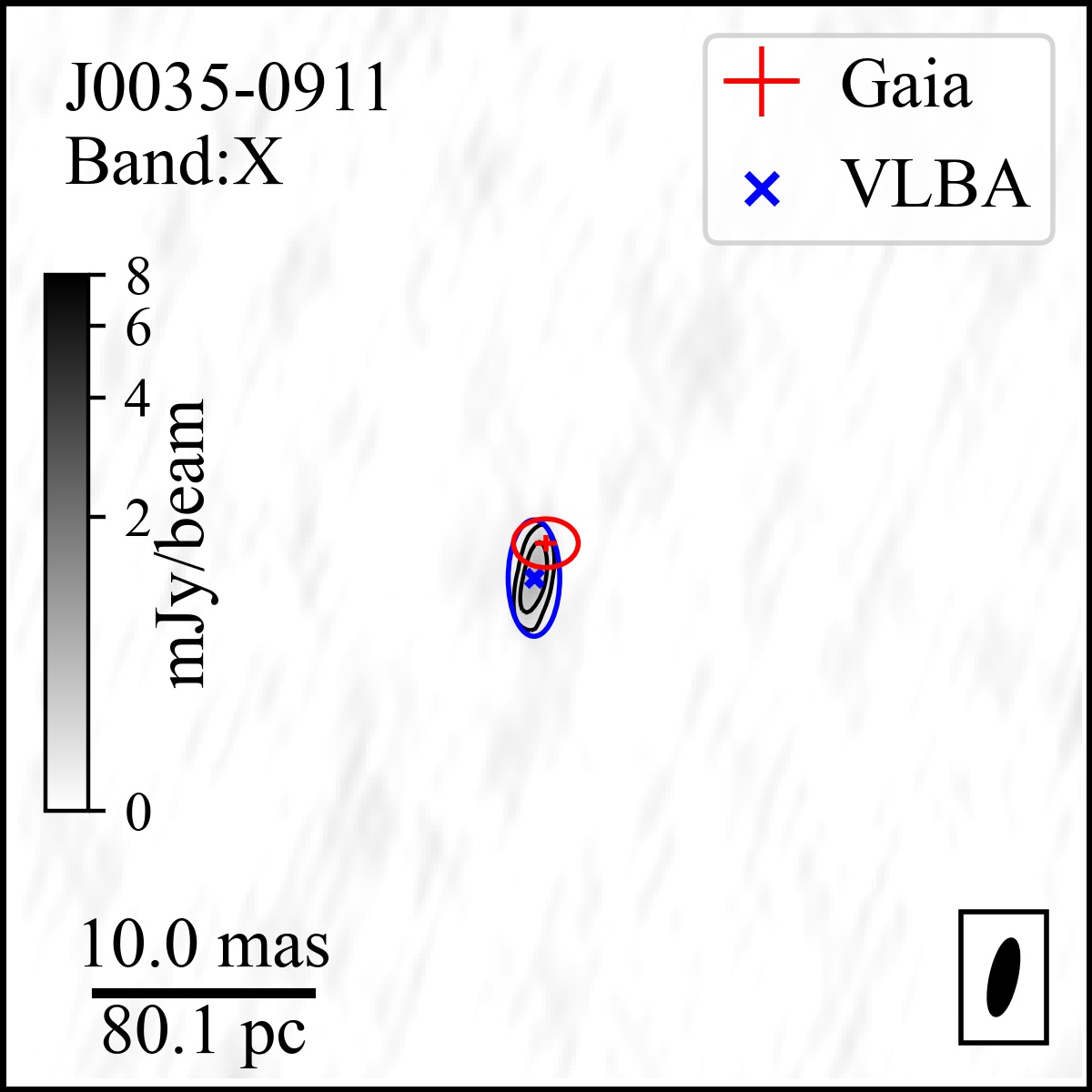}
    \fbox{\includegraphics[width=0.22\textwidth]{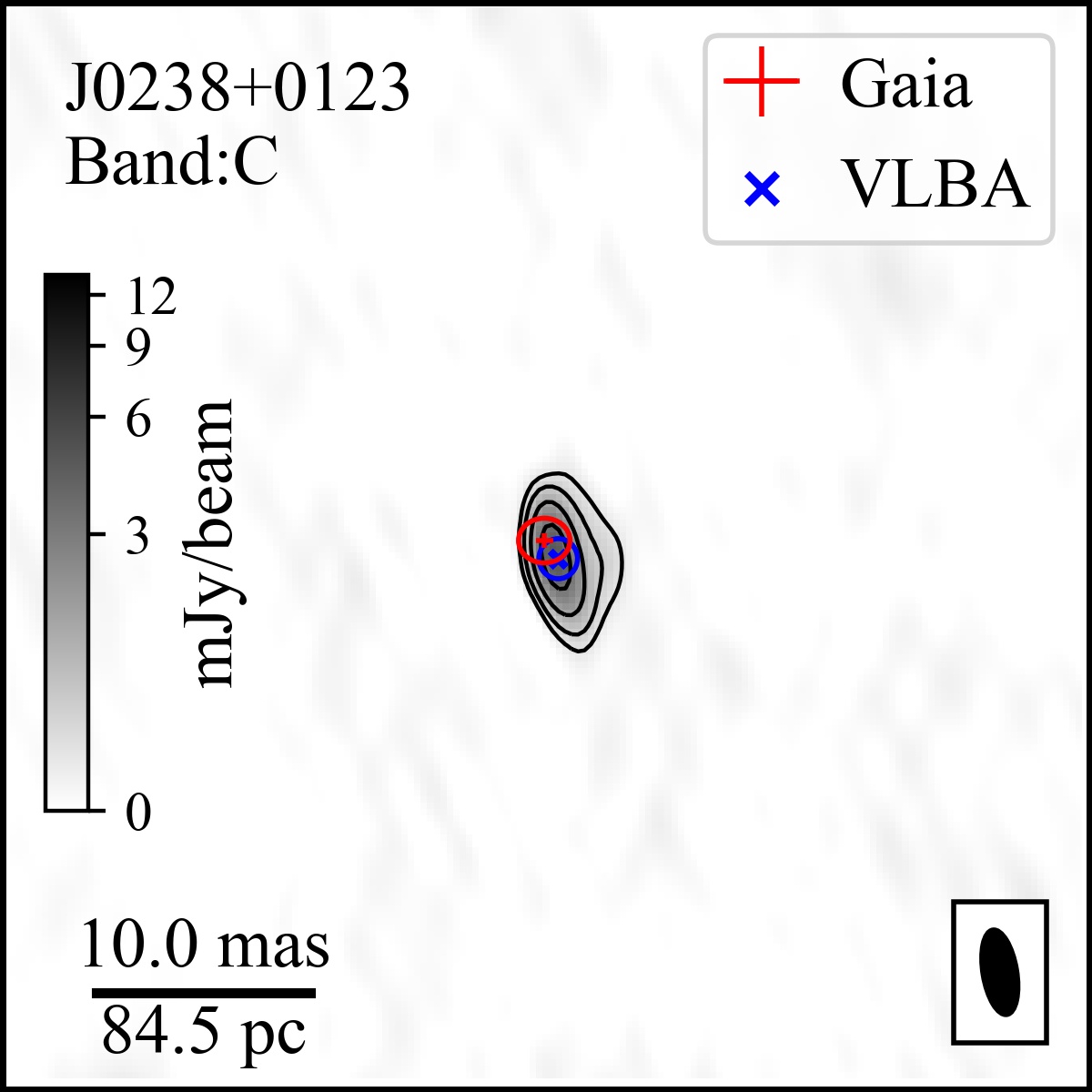}
    \includegraphics[width=0.22\textwidth]{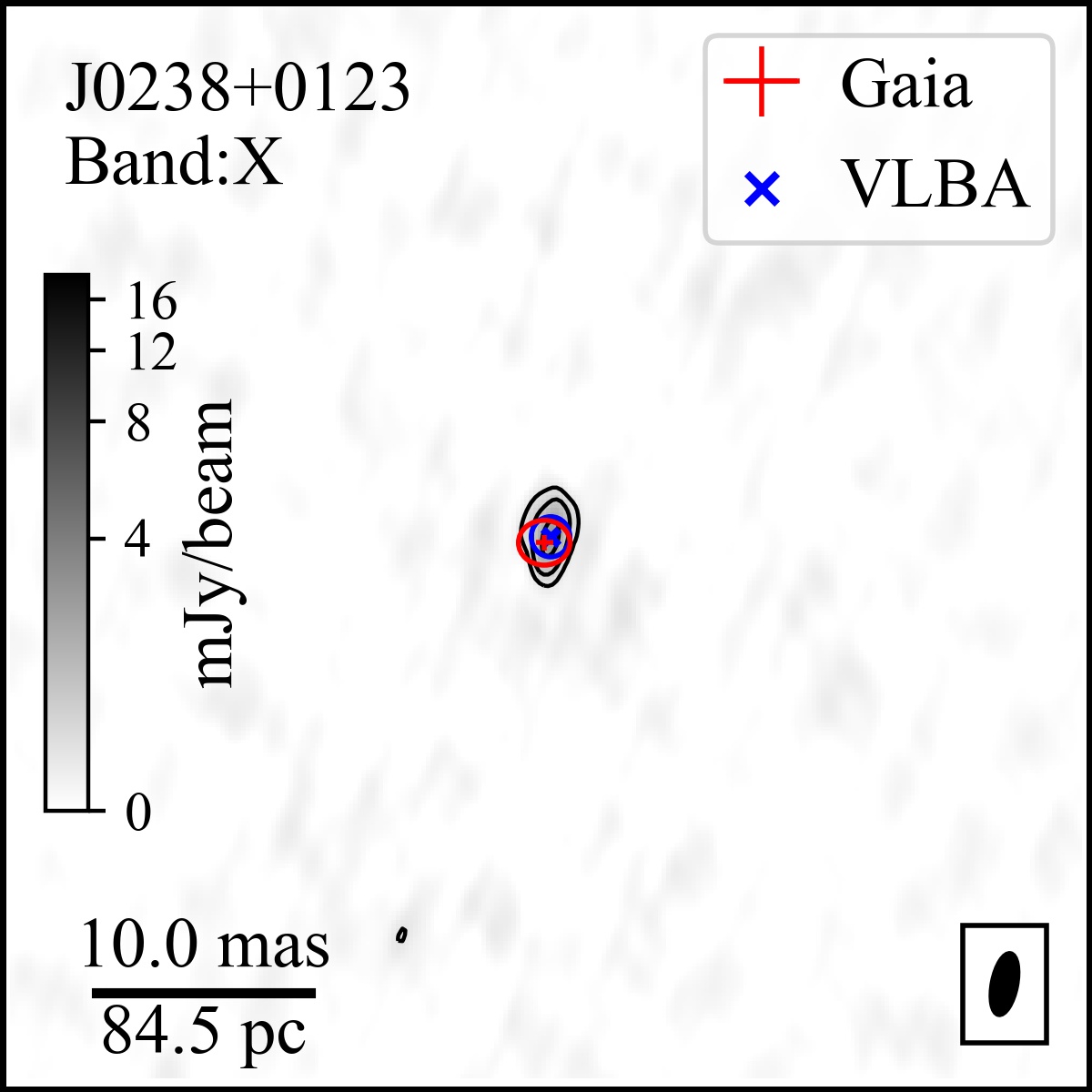}}
    \includegraphics[width=0.22\textwidth]{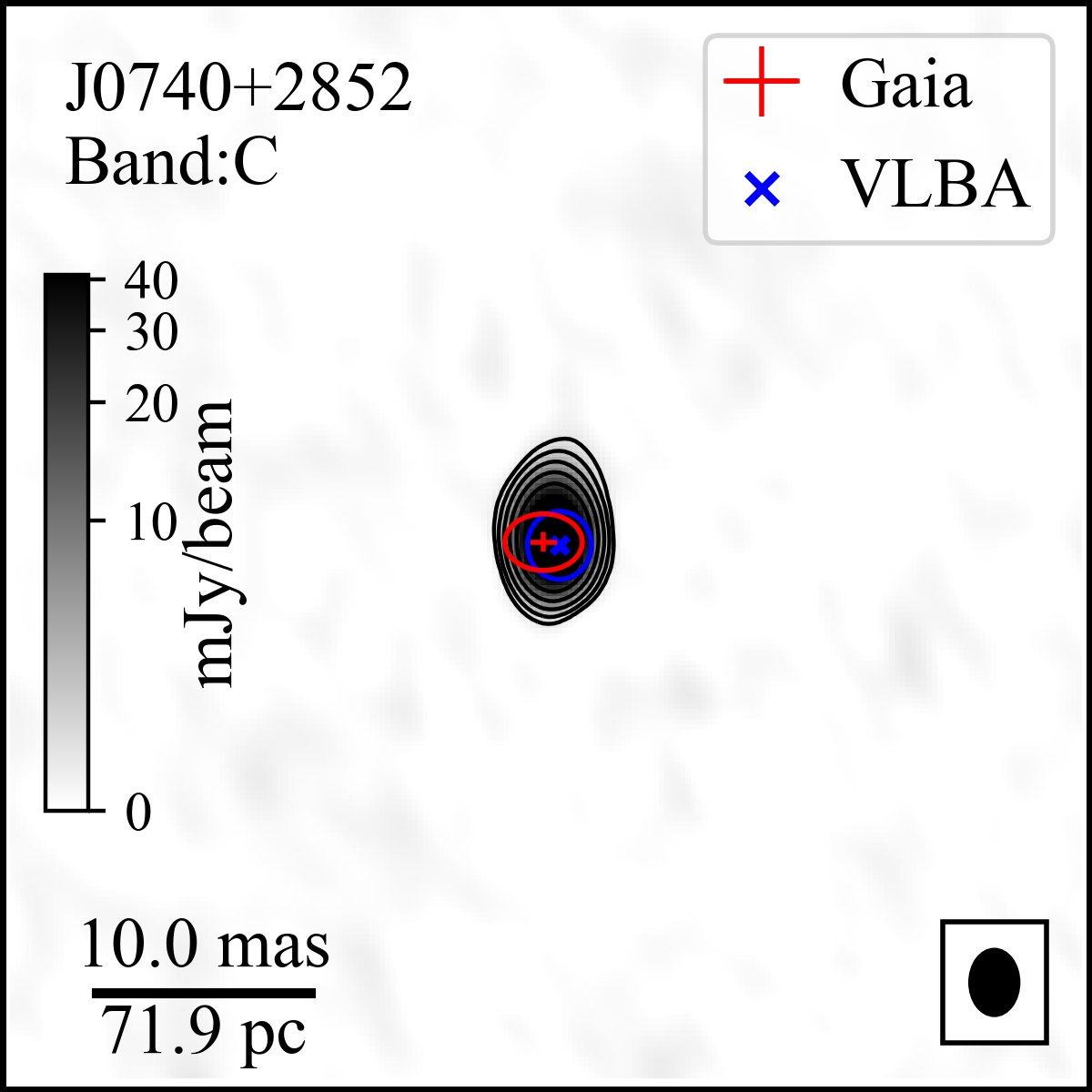}
    \includegraphics[width=0.22\textwidth]{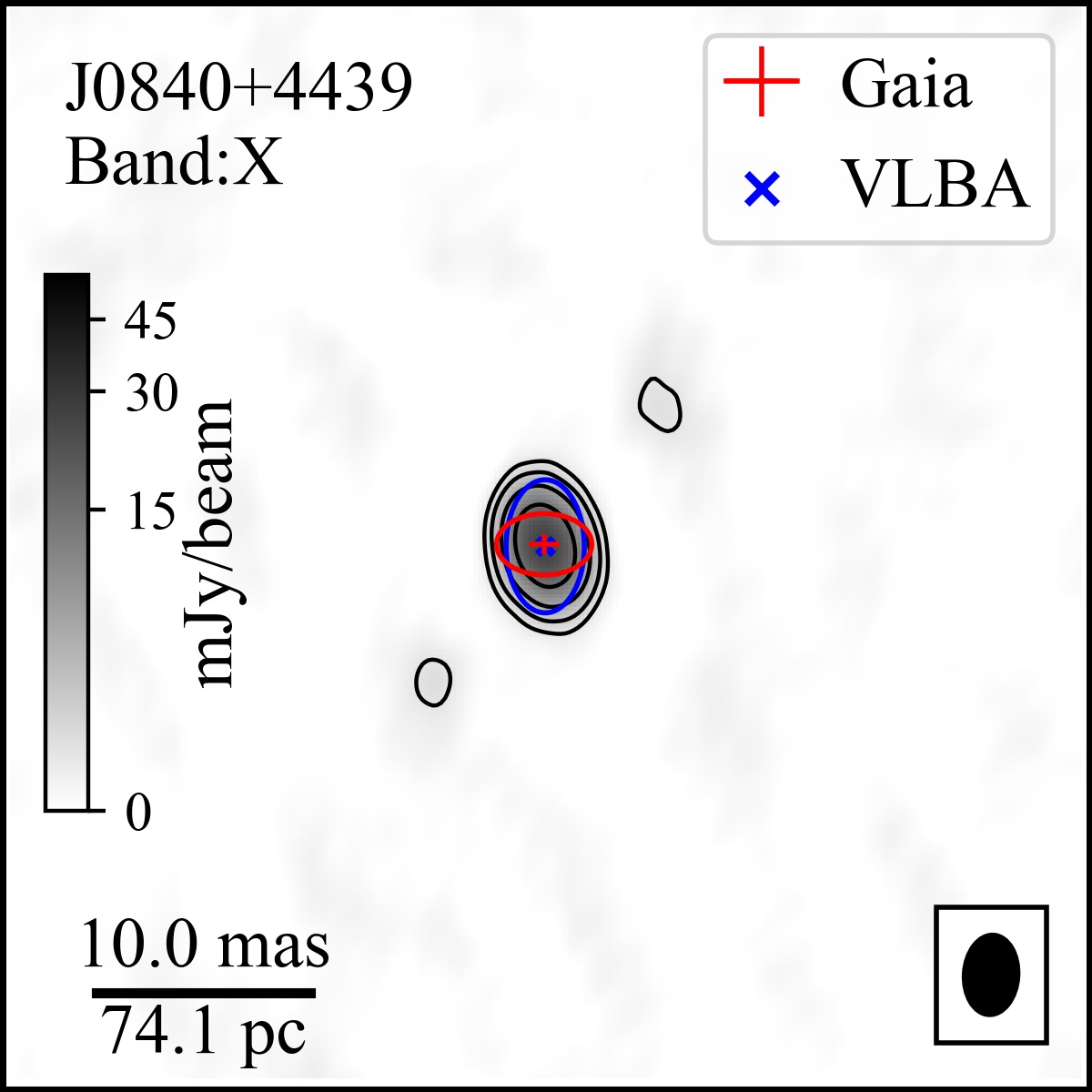}
    \includegraphics[width=0.22\textwidth]{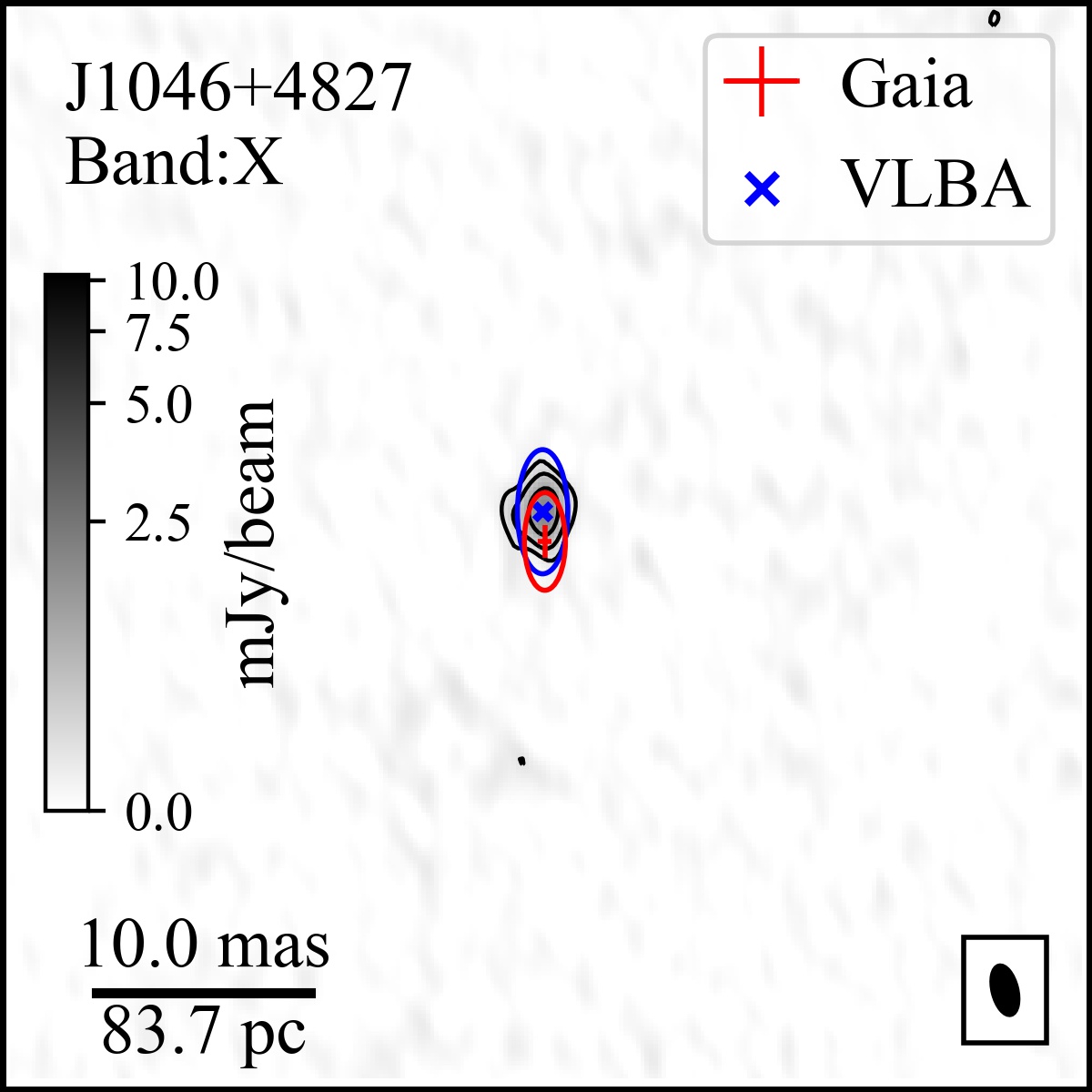}
    \includegraphics[width=0.22\textwidth]{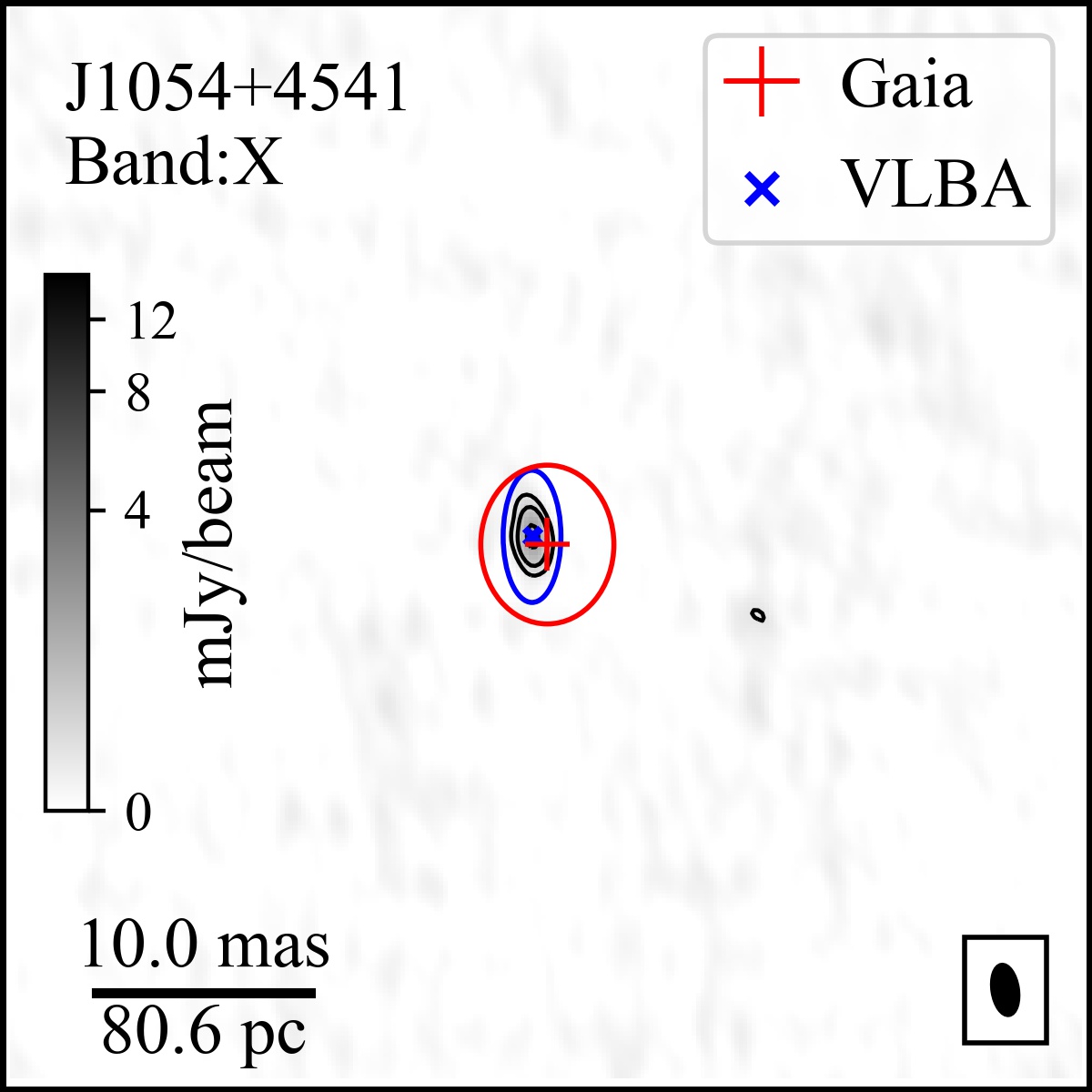}
    \includegraphics[width=0.22\textwidth]{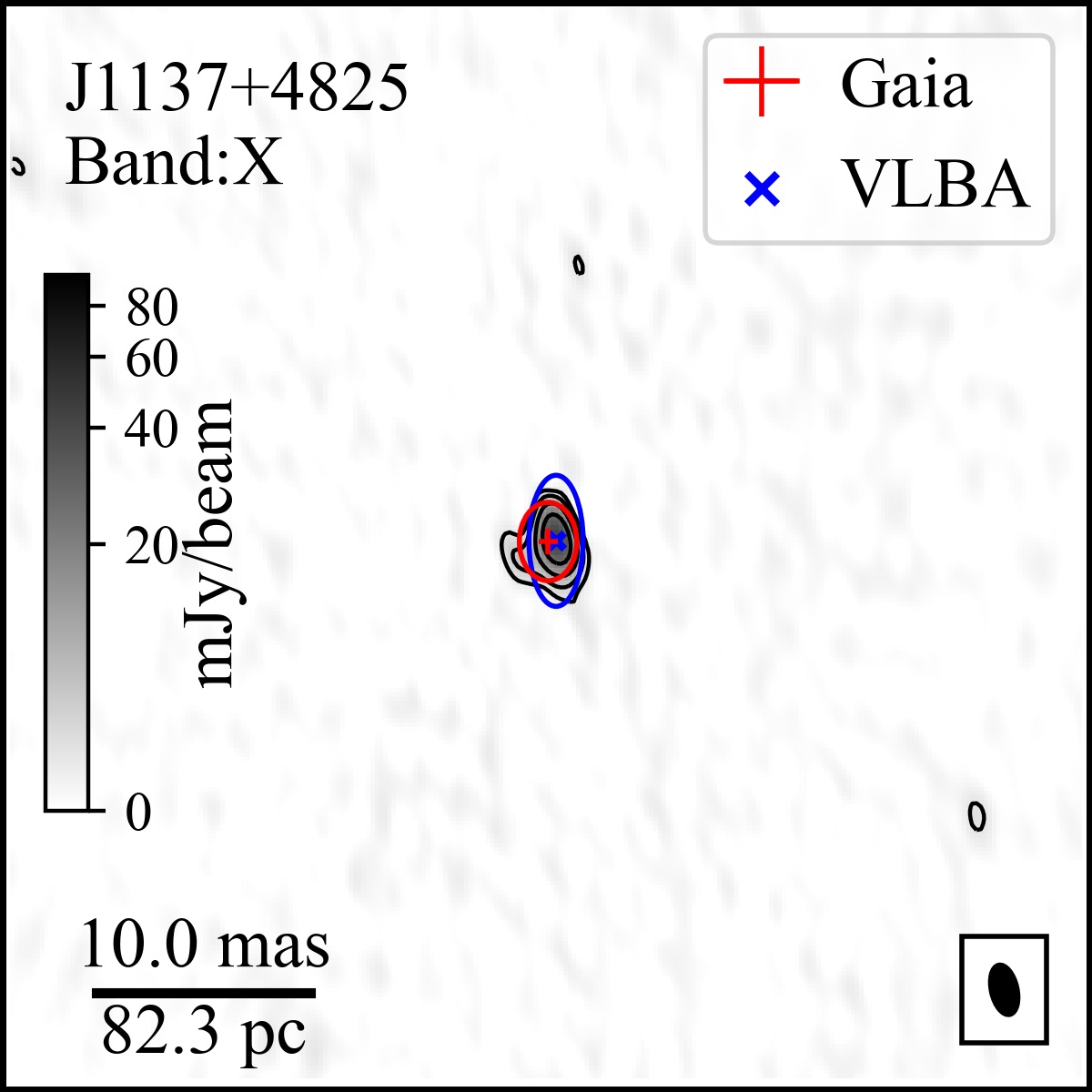}
    \includegraphics[width=0.22\textwidth]{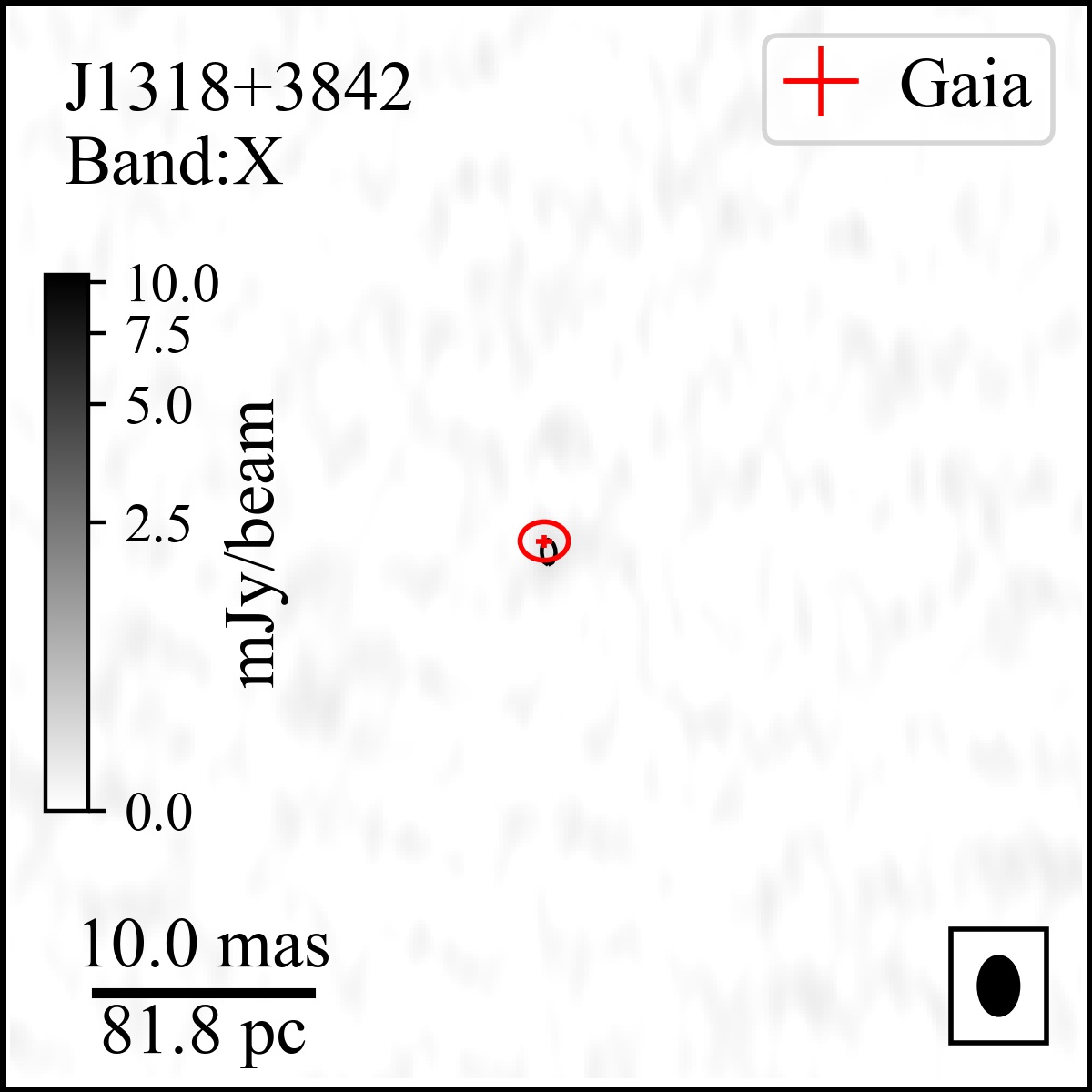}
    \includegraphics[width=0.22\textwidth]{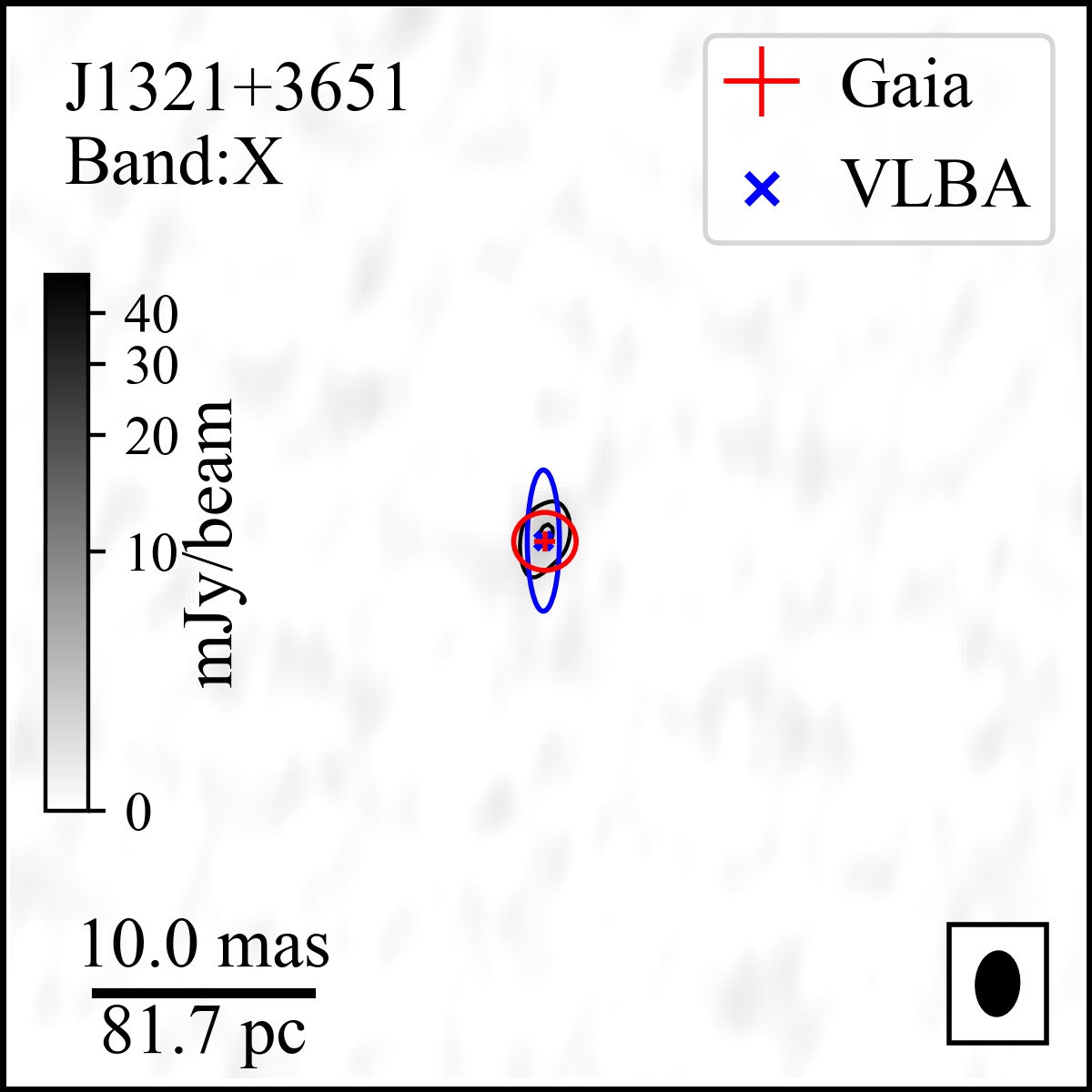}
    \includegraphics[width=0.22\textwidth]{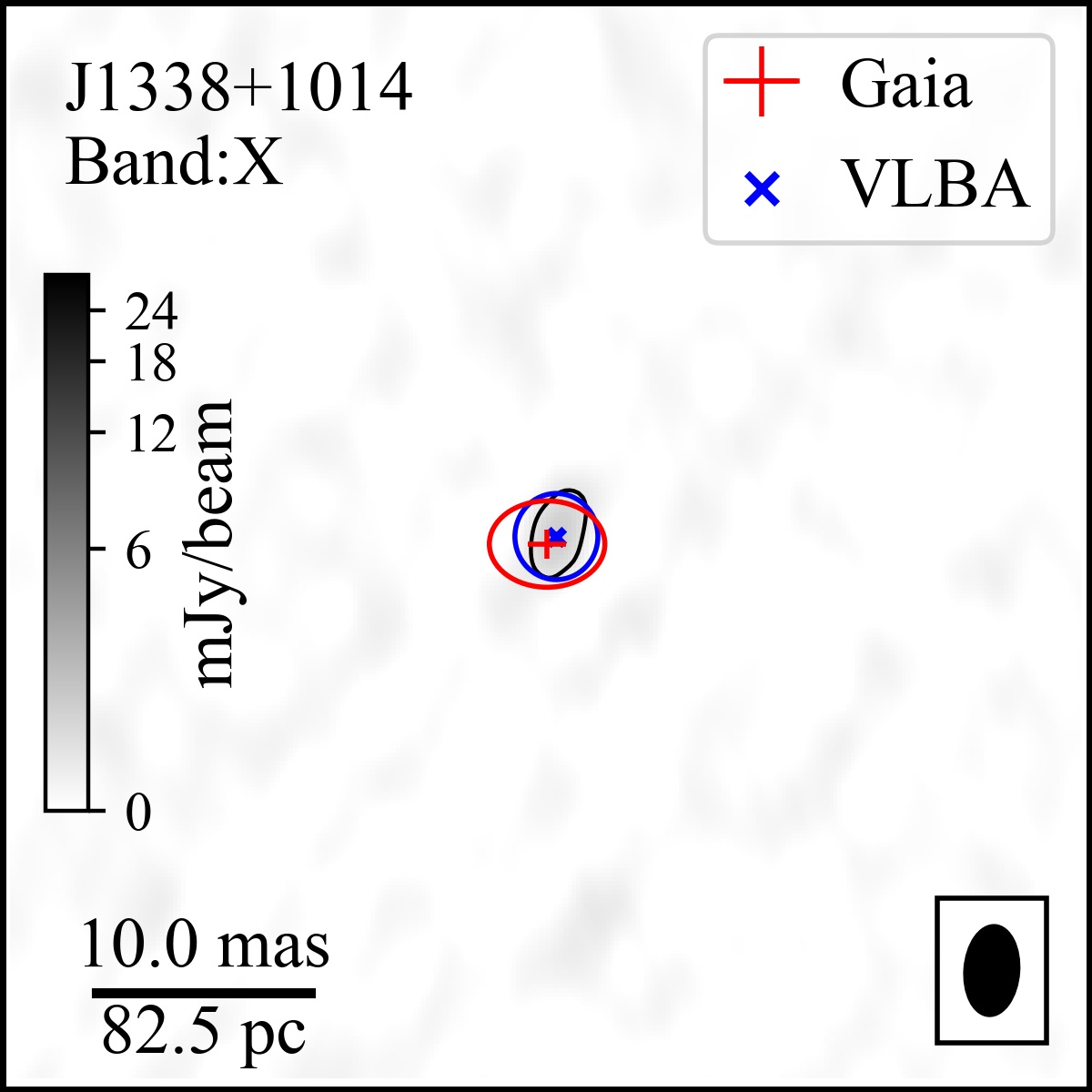}
    \includegraphics[width=0.22\textwidth]{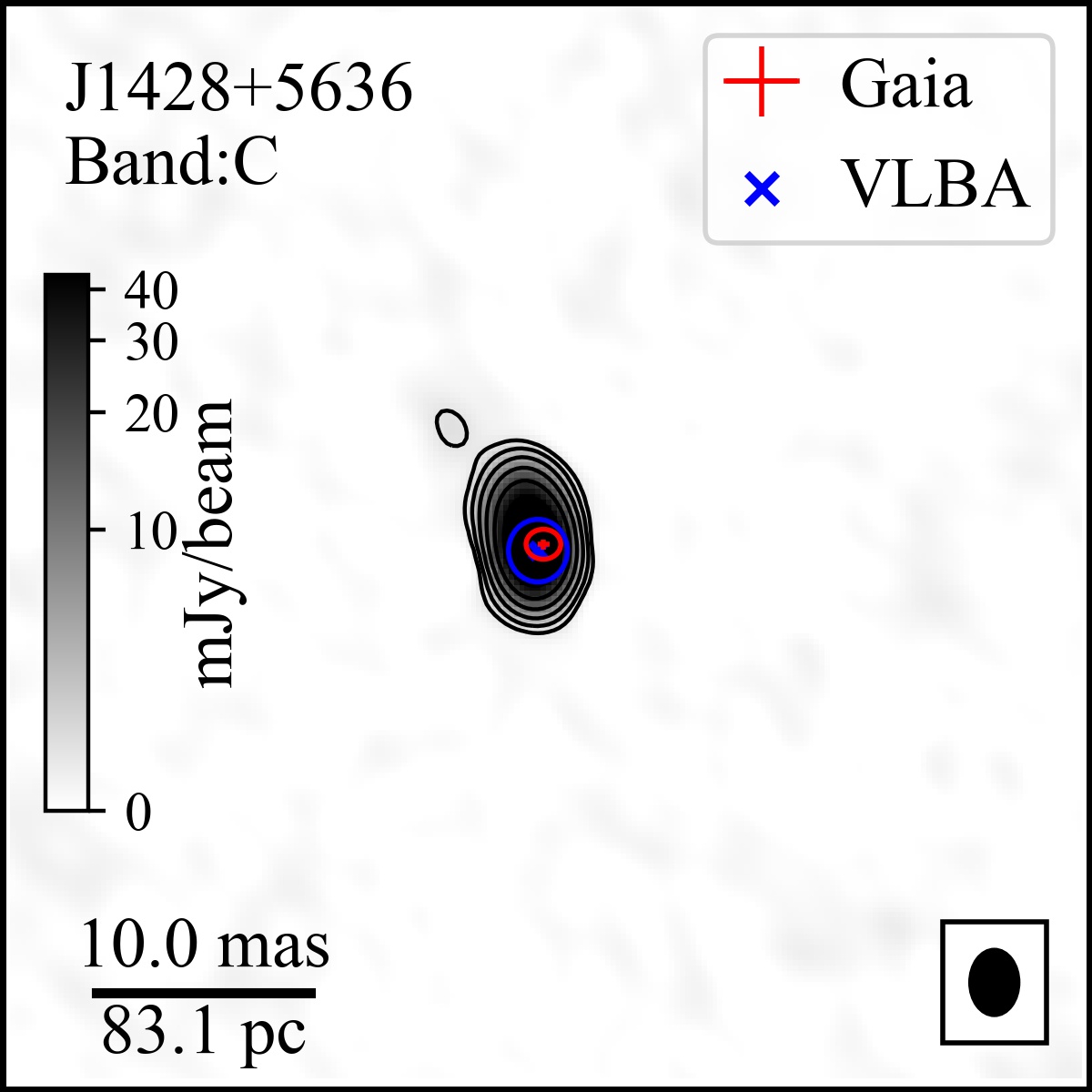}
    \includegraphics[width=0.22\textwidth]{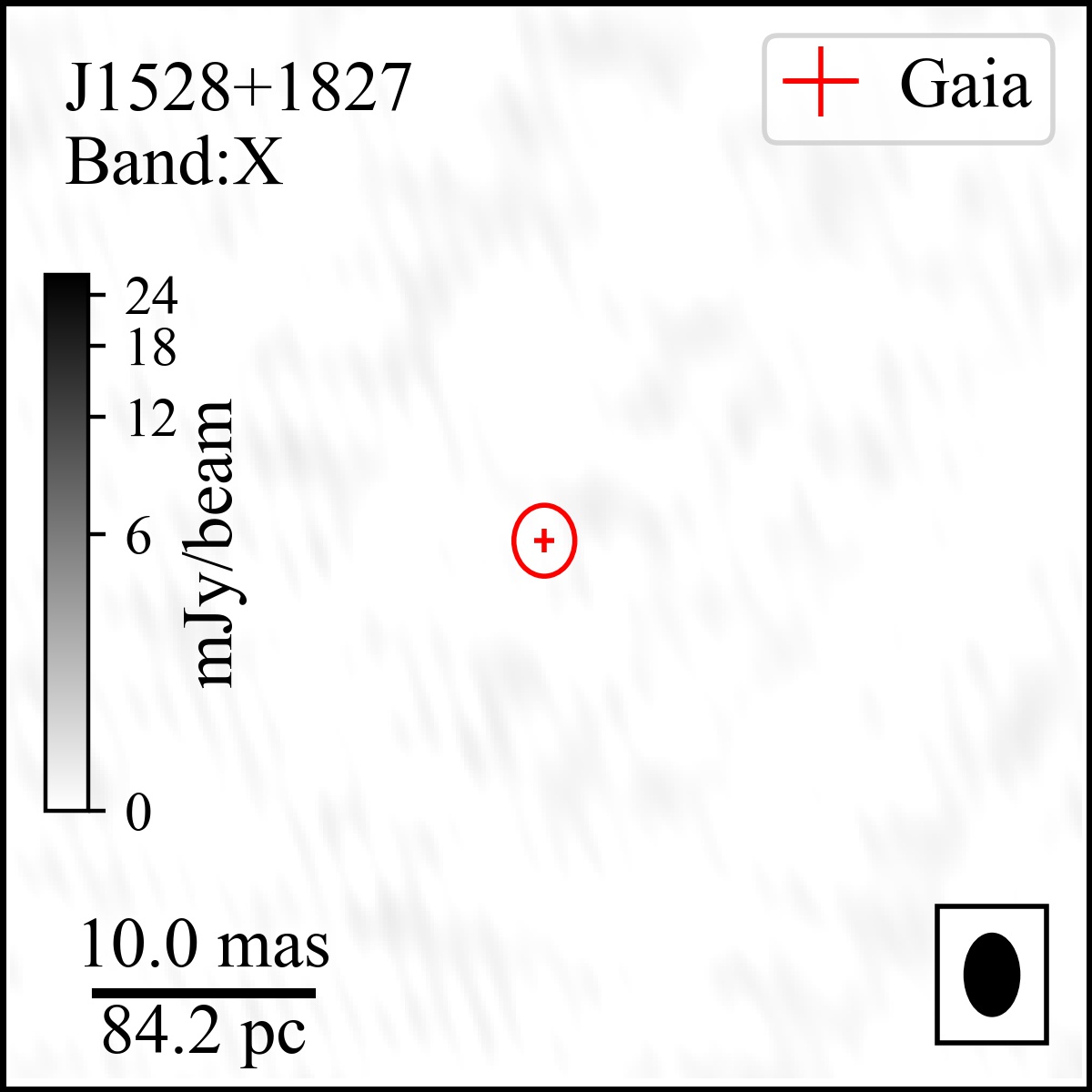}
    \includegraphics[width=0.22\textwidth]{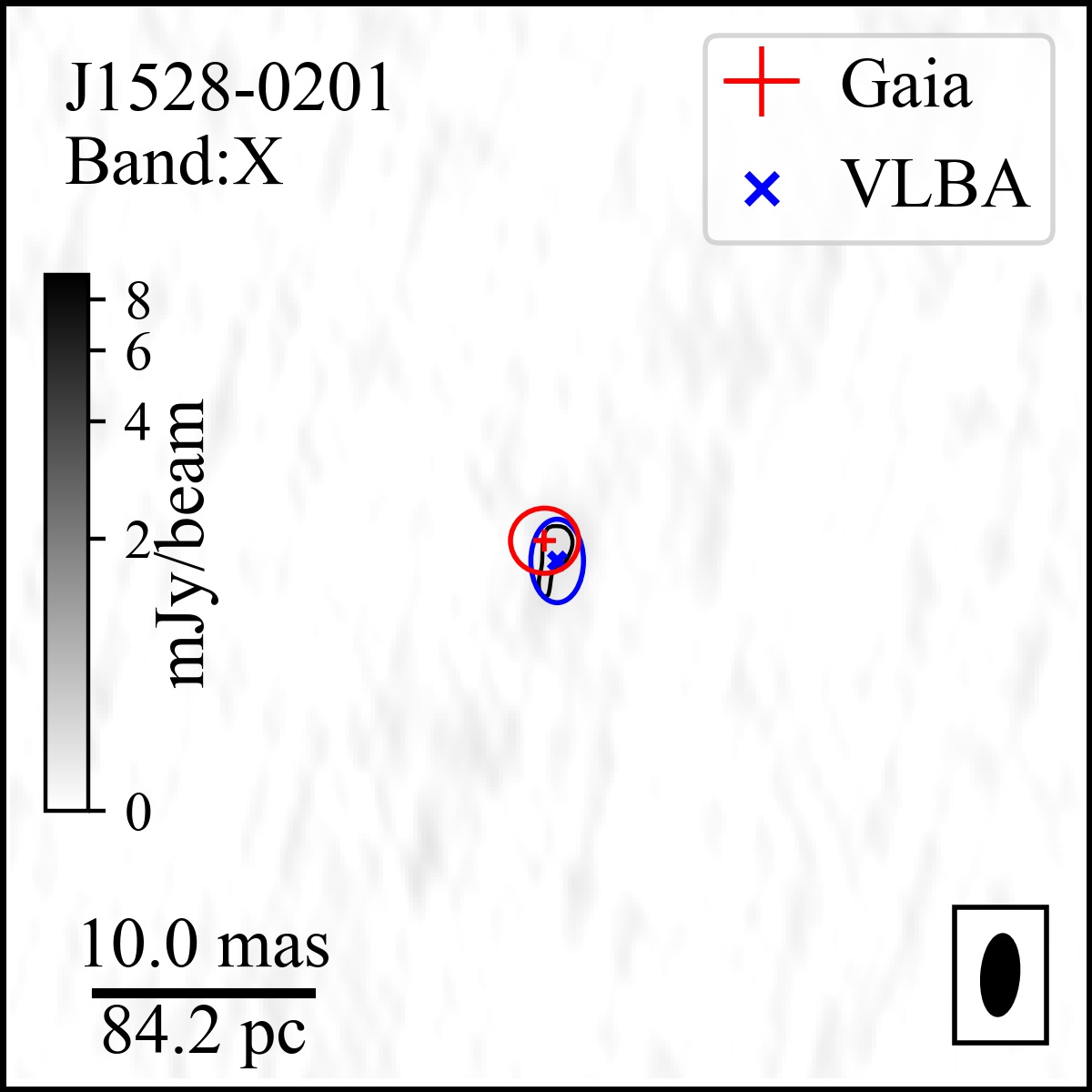}
    \includegraphics[width=0.22\textwidth]{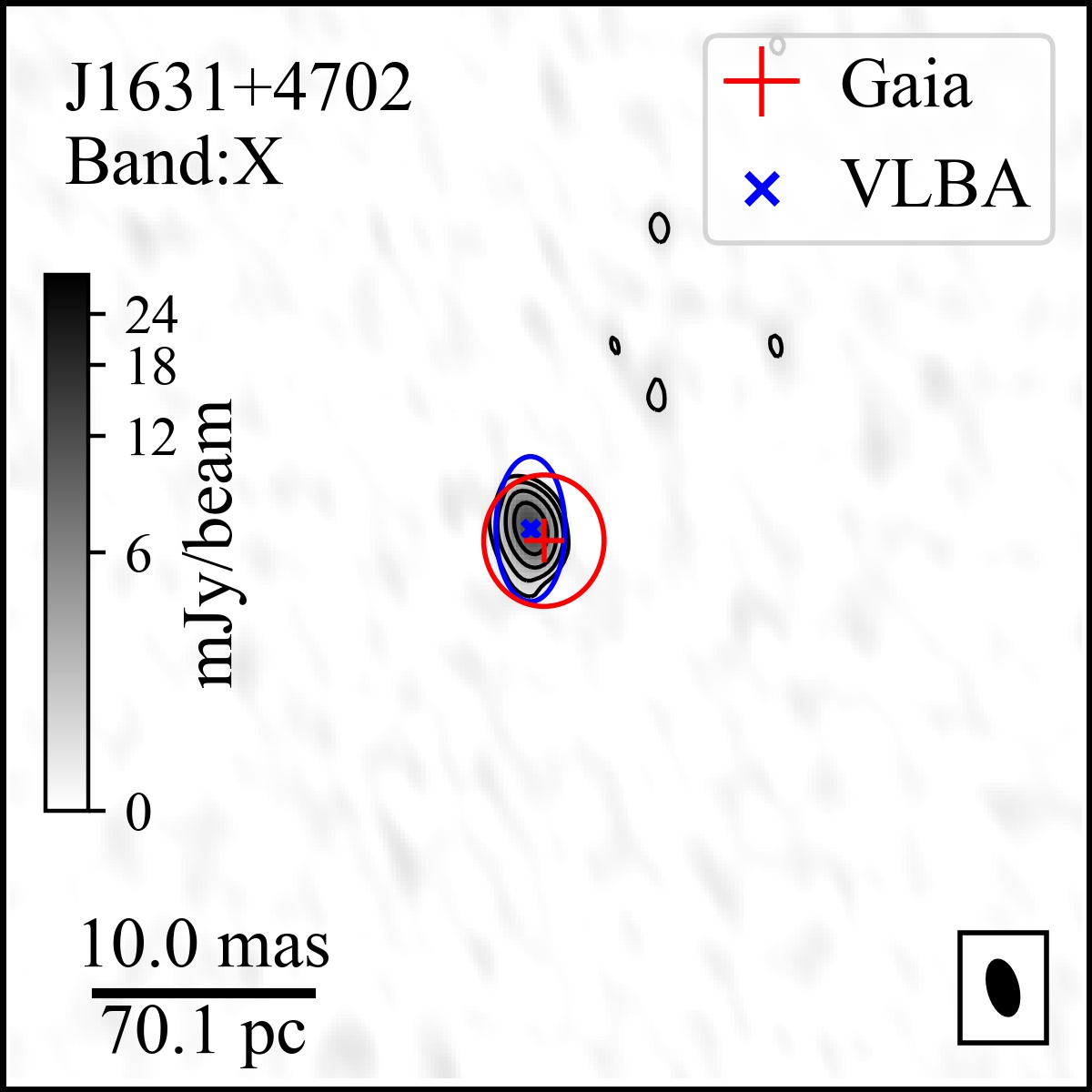}
    \fbox{\includegraphics[width=0.22\textwidth]{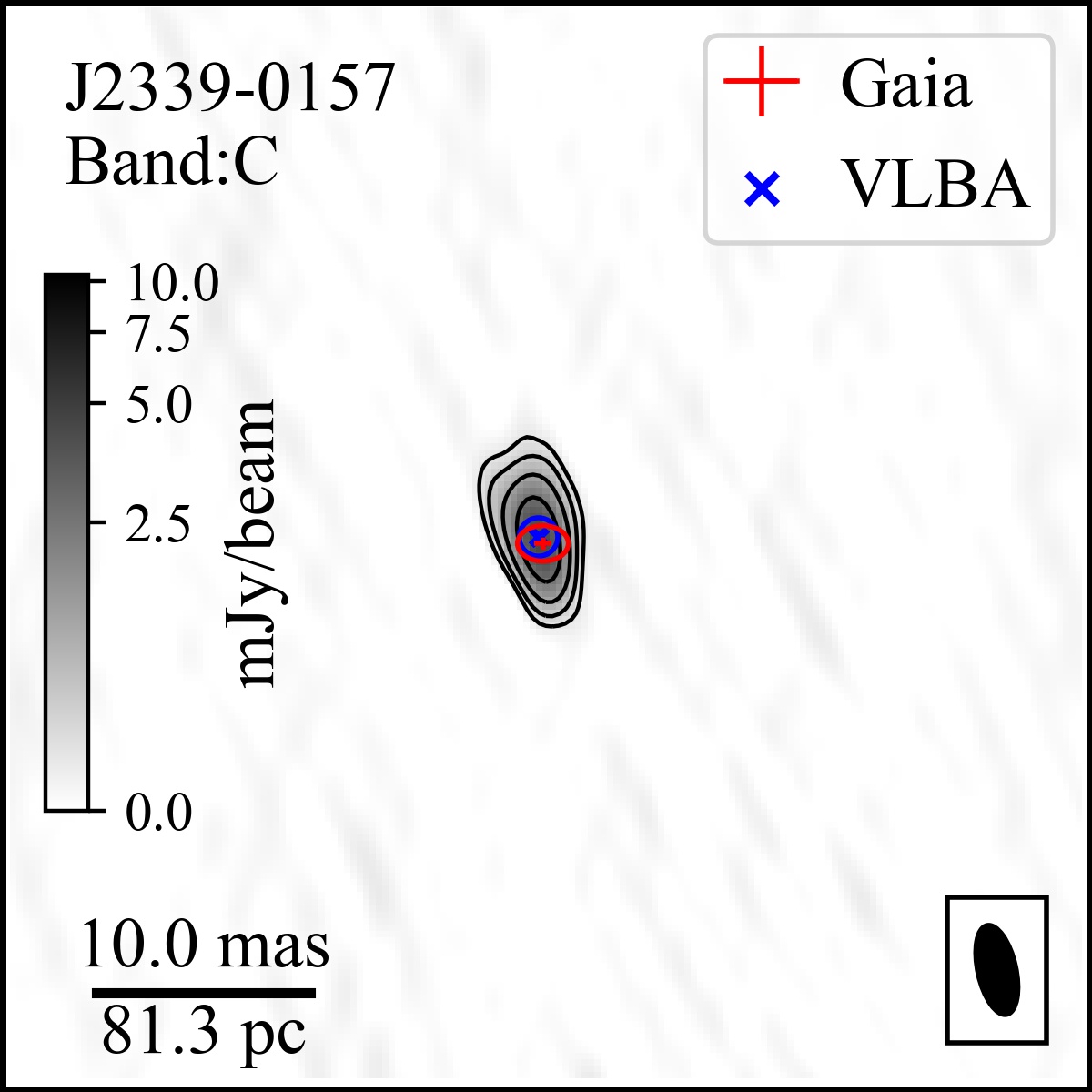}
    \includegraphics[width=0.22\textwidth]{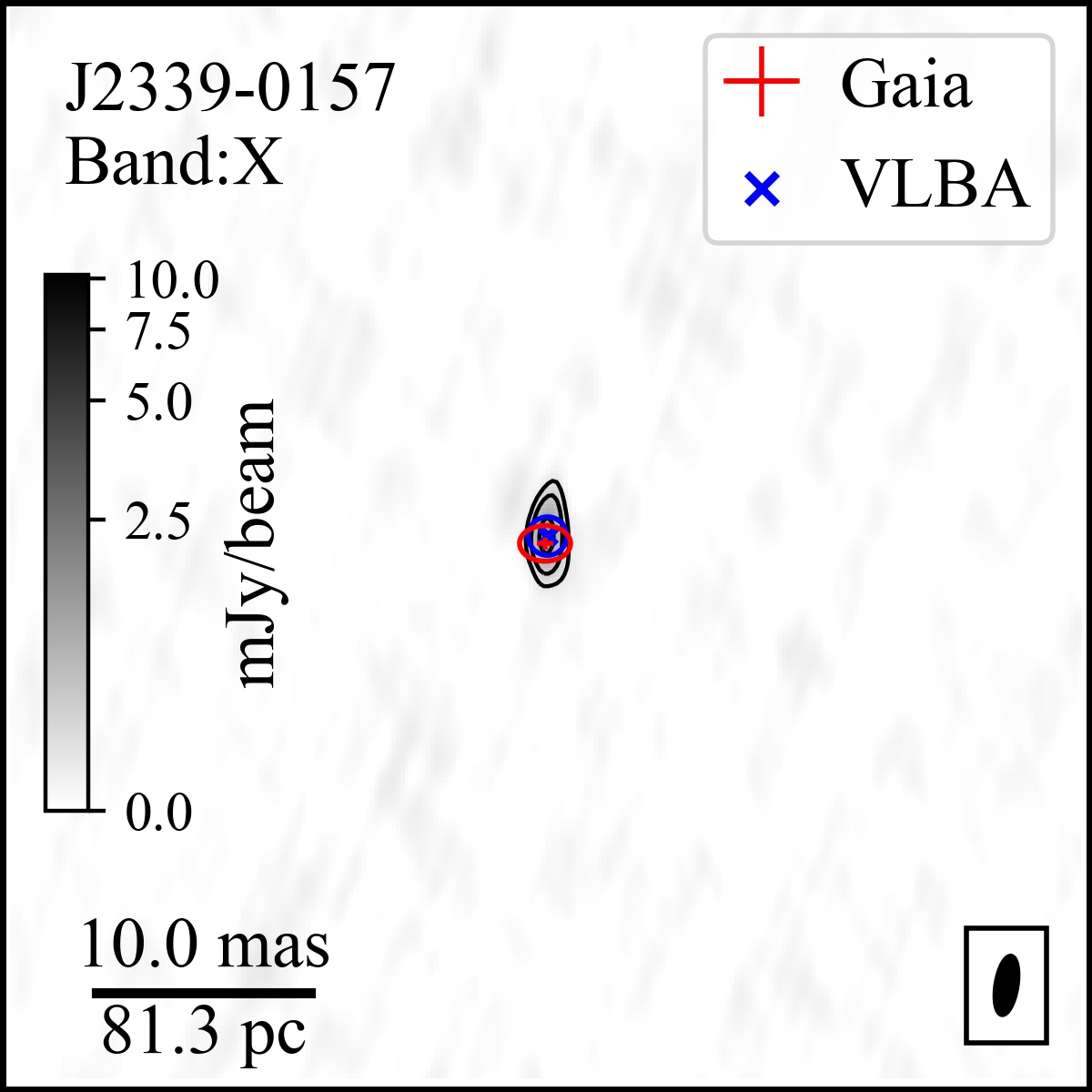}}
    
    \caption{
    VLBA (C-band and/or X-band) continuum images of the 15 targets that either show consistent positions between VLBA and Gaia, or are undetected. Other information is the same as those in \autoref{fig:VLBA_images_offset_1}}
    \label{fig:VLBA_images_no_offset}
\end{figure*}
We present the VLBA continuum images for the 23 targets. We divide these targets into two groups based on their characteristics. 
Figure \ref{fig:VLBA_images_offset_1} and Figure \ref{fig:VLBA_images_offset_2} display the VLBA images of the 8 targets that exhibit significant ($>3\sigma$) VLBA-Gaia offsets or/and have multiple radio components. Figure \ref{fig:VLBA_images_no_offset} shows the VLBA images of the 15 targets that either show consistent positions between VLBA and Gaia, or are undetected. 
The images in Figure \ref{fig:VLBA_images_offset_1}, \ref{fig:VLBA_images_offset_2}, and \ref{fig:VLBA_images_no_offset} are 50 mas $\times$ 50 mas in size and are centered at the Gaia positions. However, for J1557+1236, we used a larger image size of 150 mas $\times$ 150 mas to accommodate the VLBA position.
The majority of the images have a typical 1-$\sigma$ noise level of approximately 0.1~mJy~beam${}^{-1}$. However, for the brighter sources (such as J0749+2255, J1044+2959, and J1137+4825), the 1-$\sigma$ noise levels reach as high as approximately 1~mJy~beam${}^{-1}$. Further information about the individual images can be found in Table~\ref{tab:fitting}.

\subsection{Source Detections and Properties}\label{sec:sources}

Of the 18 targets observed in our new VLBA observations, 16 are detected with a 5-$\sigma$ detection threshold, and 2 are undetected. We search for sources within a 0\farcs6 radius and model them with 2-dimensional Gaussian components using {\sc jmfit} in AIPS. Each image is fit using a single Gaussian component. For sources with extended emission or sub-structures, we fit only a single Gaussian to the brightest component. Most of targets exhibit single compact components, while a few targets (J1051+2119, J1110+4817 and J1259+5140) show second compact components or extended structures at the scales of dozens mas. We do not find any compact source at large scale ($>$100 mas), except for the known dual quasar J0749+2255, which has a separation of 0\farcs46 \citep{Shen2021,ChenYC2023}. The VLBA radio measurements and image properties for each target are summarized in Table \ref{tab:fitting}. Previous analyses have suggested that differences in the $u$-$v$ coverage of the observations may contribute up to 10\% in systematic uncertainties. The coordinate positions obtained from the C-band and Ku-band images are consistent with those obtained from the X-band images, with differences of less than 0.1~mas, consistent with the astrometric accuracy. The 5-$\sigma$ upper limits of the two undetected targets, J1318+3842 and J1528+1827, are also provided in Table~\ref{tab:fitting}.

\subsection{Spectral Index}\label{sec:si}

\begin{figure}
    \centering
    \includegraphics[width=0.49\columnwidth]{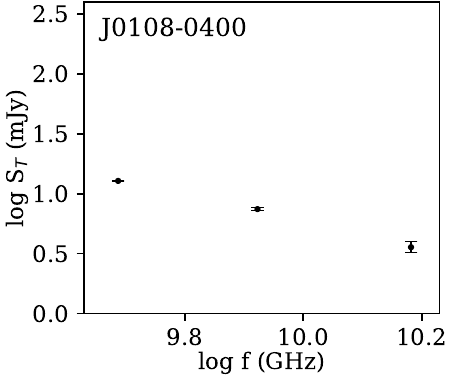}
    \includegraphics[width=0.49\columnwidth]{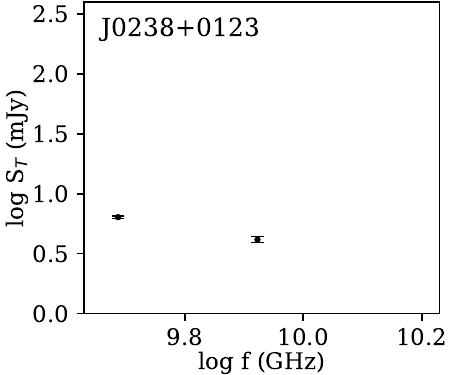}
    \includegraphics[width=0.49\columnwidth]{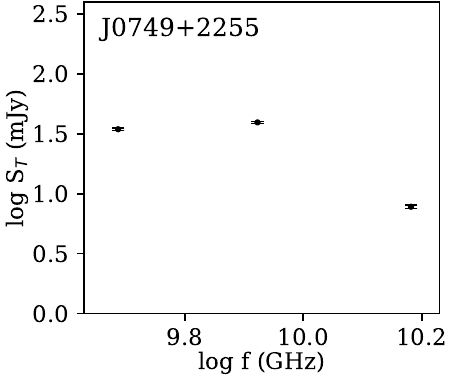}
    \includegraphics[width=0.49\columnwidth]{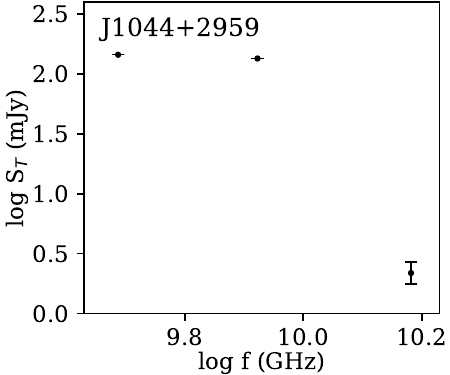}
    \includegraphics[width=0.49\columnwidth]{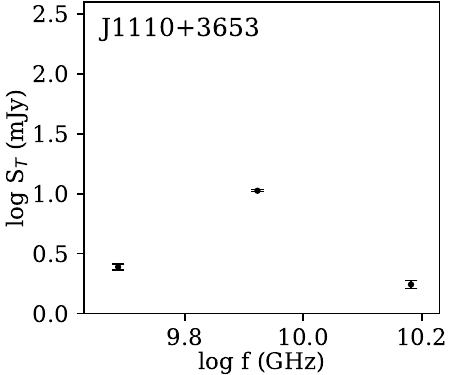}
    \includegraphics[width=0.49\columnwidth]{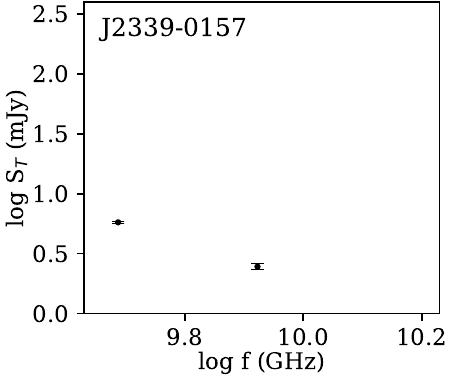}
    
    \caption{
    %\textbf{\color{red} Please add colorbars to show surface brightness (similar to your Nature paper VLA figure). Quote the typical rms noise levels (i know we list them in the table, but it would be good to be reminded about the typical sensitivity reached also directly in the figure caption). Same for Figure 4.} \textbf{\color{blue} YCC:Done}
    Radio spectra of the six targets for which we have detections at at least two frequencies.}
\label{fig:radio_spectra}
\end{figure}

We have determined the spectral indices of the six sources for which we have detections at at least two frequencies. These six targets comprise the four targets with significant VLBA-Gaia offsets and the two targets with extended structures. The spectral index, denoted as $\alpha$, is defined as S$_{\nu} \propto \nu^{\alpha}$, where S$_{\nu}$ is the radio flux density and $\nu$ is the frequency. We present the spectral indices (Table \ref{tab:spectral_index}) and the radio spectra (Figure \ref{fig:radio_spectra}) for the six targets. 
%\ref{tab:spectral_index} presents the spectral indices of the six targets and Figure \ref{fig:radio_spectra} show the radio spectra of the six targets.
Three targets, namely J0749+2255, J1044+2959, and J1110+3653, exhibit flat or rising spectra with $\alpha\gtrsim0$, suggesting synchrotron self-absorbed cores. The remaining three targets have steep spectra with $\alpha\lesssim-1$, indicating optically-thin radio jets. The spectral indices and radio emission classifications of these targets are also listed in Table \ref{tab:spectral_index}. 

\begin{deluxetable}{ccccc}
 \tablecaption{Spectral indices, morphology, and classifications of the radio emission for the targets that have multi-band observations, exhibit significant VLBA-Gaia offsets, or show multiple radio components.
 \label{tab:spectral_index}}
 \tablehead{\colhead{Name} & \colhead{$\alpha_{\rm C-X}$} & \colhead{$\alpha_{\rm X-Ku}$} & 
 \colhead{Classification} & 
 \colhead{Note} \\
     \colhead{(1)} & 
     \colhead{(2)} & 
     \colhead{(3)} & 
     \colhead{(4)} &
    \colhead{(5)} 
 }
 \startdata
 J2339$-$0157 & -1.57 & \nodata & jet & \nodata \\
 J0108$-$0400 & -1.00 & -1.23 & jet & offset\\
 J0238+0123 & -0.80 & \nodata & jet & \nodata \\
 J0749+2255 & 0.24 & -2.72 & core & offset\\
 J1044+2959 & -0.13 & -6.92 & core & offset\\
 J1051+2119 & \nodata & \nodata & jet & multiple\\
 J1110+3653 & 2.71 & -3.02 & core & offset\\
 J1110+4817 & \nodata & \nodata & jet & offset,multiple\\
 J1259+5140 & \nodata & \nodata & jet & multiple\\
 J1557+1236 & \nodata & \nodata & unknown & offset\\
 \enddata
 \tablecomments{Column 1: Name. Column 2: Spectral index using C-band and X-band flux. Column 3: Spectral index using X-band and Ku-band flux. Column 4: Classification based on morphology and spectral index. Column 5: ``offset": VLBA-Gaia offset. ``multiple": multiple radio components.}
\end{deluxetable}

\subsection{VLBA-Gaia offset}
\label{gaia_vlba_offset}

\begin{figure}
    \centering    \includegraphics[width=0.5\textwidth]{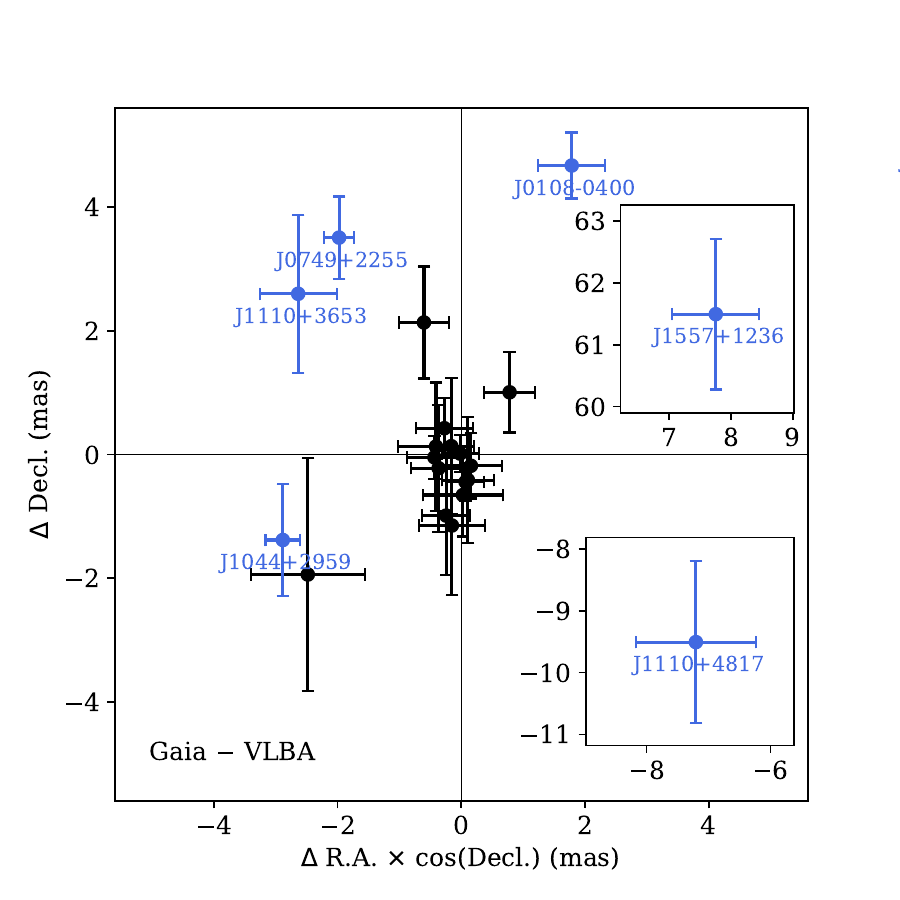}
    \caption{Distribution of positional offsets between the Gaia DR3 optical position and the VLBA radio source positions for the 21 VLBA-detected targets. Targets with offsets greater than $3\sigma$ are highlighted in blue.
    The error bars represent $1\sigma$ errors considering the fitting error, the phase calibrator's position uncertainty, and the error associated with the phase-referencing technique, as described in Section \ref{gaia_vlba_offset}. The axes of the insets are in units of mas, identical to those in the the main plot.}
    \label{fig:offset_gaia_vlba}
\end{figure}

We compare the optical positions from Gaia DR3 with the VLBA radio positions to identify any potential dual, offset, or lensed quasar candidates. The VLBA-Gaia offsets are calculated based on the X-band images. The VLBA position uncertainty is determined using the quadrature sum of the fitting error, the phase calibrator's position uncertainty, and the error associated with the phase-referencing technique \citep{Pradel2006}. We obtain the phase calibrator positions from the VLBA calibrator list \citep{Charlot2020, Petrov2021}. The final VLBA position errors were within the range of 0.3 to 1.2 mas. Figure \ref{fig:offset_gaia_vlba} displays the positional offsets' distribution for the 21 targets detected by the VLBA. Among the VLBA-detected targets, 6 have radio counterparts significantly offset from the Gaia positions using the 3-$\sigma$ threshold. Further discussion of the origin of radio emission for these VLBA-Gaia offset targets is presented in Section \ref{individual_cases}.

\section{Discussion}\label{sec:discussions}

\subsection{Targets with multiple radio detections or significant VLBA-Gaia offsets}\label{individual_cases}

Out of the 16 detected targets, 8 targets exhibit either multiple radio detections or significant VLBA-Gaia offsets (Figure \ref{fig:VLBA_images_offset_1} and \ref{fig:VLBA_images_offset_2}). 
Based on the spectral indices and morphology, we examine the potential origins of the radio emission for the 8 targets. If a target shows two flat-spectrum cores or a flat-spectrum core with significant Gaia-VLBA offset, we classify it as a dual quasar candidate. If a target shows steep spectral index or extended structure along the Gaia-VLBA offset, the radio emission is likely from a jet.

\subsubsection{J0108$-$0400}

J0108$-$0400 exhibits a 5.0$\pm$0.5 mas (corresponding to 42 pc) VLBA-Gaia offset. The steep spectral index ($\alpha_{C-X}=-1.57$) of the radio component and its extension toward the Gaia position suggest the presence of a diffuse lobe structure. Therefore, the offset is most likely caused by a small-scale radio jet.

\subsubsection{J0749+2255}
J0749+2255 shows a 4.0$\pm$0.6~mas (corresponding to~33~pc) VLBA-Gaia offset. The radio component has a inverted spectral index of 0.24, indicating a self-absorbed core. J0749+2255 was already identified as a 0\farcs46 dual quasar \citep{ChenYC2022}, with both nuclei detected in our VLBA images and reported in \citet{Shen2021}.  
Gaia only reports one detection for J0749+2255, so the Gaia optical position may be affected by the blended second nucleus, leading to the offset. Nonetheless, the possibility of a third SMBH at a pc scale centered at the optical component cannot be ruled out.

\subsubsection{J1044+2959}
J1044+2959 shows a 3.2$\pm$0.5 mas (corresponding to~25~pc) VLBA-Gaia offset. The radio component is compact and has a flat spectral index of $-$0.13, suggesting a self-absorbed core. J1044+2959 is likely a candidate dual quasar, which could explain the offset if one nucleus is radio-bright but the other is optically bright (Panel~b, Figure~\ref{fig:cartoon}). Alternatively, both quasars could be optically bright, but only one is radio-bright (Panel~a,  Figure~\ref{fig:cartoon}), resulting in an offset between the light centroid of the total optical flux from the radio centroid.

\subsubsection{J1051+2119}
J1051+2119 shows a consistent position between VLBA and Gaia, but it presents a second radio component located toward the east. The second component shows an elongated structure connecting to the primary component, which suggests that it is a small-scale jet. To confirm the jet scenario, a spectral index analysis derived from multi-band images is necessary, though we only have an archival C-band image.

\subsubsection{J1110+3653}
J1110+3653 exhibits a 3.7$\pm$1.0 mas (corresponding to 25 pc) VLBA-Gaia offset. The radio component is compact and has a inverted spectrum ($\alpha_{\rm C-X}=2.71$ and $\alpha_{\rm X-Ku}=-3.02$), consistent with a self-absorbed core. Similar to the discussion for J1044+2959, J1110+3653 is likely a candidate dual quasar.  In addition, given its low redshift of 0.63, J1110+3653 could be a candidate off-nucleus quasar where the optical center is shifted because of emission from the host galaxy (Panel c in Figure \ref{fig:cartoon}).

\subsubsection{J1110+4817}
J1110+4817 exhibits an 11.9$\pm$1.0 mas (corresponding to 87 pc) VLBA-Gaia offset. 
The C-band archival image reveals a possible bipolar jet that appears to originate from the Gaia position. Although the cause of the offset is likely due to the small-scale radio jet, confirmation of the jet scenario requires spectral index analysis derived from multi-band images.

\subsubsection{J1259+5140}
J1259+5140 shows a consistent position between VLBA and Gaia, but it presents several radio components positioned towards the west. The configuration of these components aligns with the direction of the small VLBA-Gaia offsets, indicating that they may constitute a small-scale jet. Nevertheless, confirming the jet hypothesis requires spectral index analysis derived from multi-band images, and we only have an archival C-band image.

\subsubsection{J1557+1236}

J1557+1236 exhibits a significant VLBA-Gaia offset of 62.0$\pm$1.2 mas (corresponding to 485 pc). The radio component is a weak detection at around 6$\sigma$ and shows slight extension. Without deeper images or additional images in other bands, the origin of the offset is unclear.

\subsection{Other possible origins of VLBA-Gaia offsets}

In Section \ref{individual_cases}, we discuss individual cases that have significant VLBA-Gaia offsets or multiple radio components assuming the origins are dual/off-nucleus quasars and small-scale radio jets. Nevertheless, under a few other possible scenarios, single normal quasars could exhibit offsets between optical and radio positions. In the following sections, we explore these scenarios.

\subsubsection{Optical Jet}
The presence of optical jets at scales of hundreds of milli-arcseconds is considered to be one of the possible mechanisms for the VLBA-Gaia offsets \citep{Petrov2017,Petrov2019}. These optical jets can cause a shift in the Gaia positions along the direction of the jet, away from the nucleus. To investigate the possibility of optical jets, we examine the optical spectra from SDSS and the optical images from the Dark Energy Camera Legacy Survey \citep[DECaLS,][]{DECaLS}, which has an image quality of approximately 1 arcsecond. We look for any indications of optical jets that could contaminate the optical spectra with a featureless power-law continuum from synchrotron radiation. Among all eight candidates, we find that they show typical spectra of type 1 broad-line quasars. Given the typical radio flux density of our sample($\sim$10 mJy), we do not expect any optical synchrotron to contribute \citep{Collinge2005}. Additionally, based on the DECaLS images, we observe that seven out of the eight candidates, excluding J0749+2255, which has already been identified as a dual quasar \citep{Shen2021,ChenYC2023}, show a point-like morphology without any signs of extended optical jets. The point-like morphology suggests that the presence of strong optical jets at scales of arcseconds is unlikely. However, faint sub-arcsecond optical jets that are unresolved in DECaLS are still possible.

\subsubsection{Extended Host Galaxy}

The astrometric solutions of Gaia detections can be affected by optical emission from host galaxies. According to a systematic study by \citet{Makarov2019}, the fraction of sources with VLBA-Gaia offsets is notably higher at low redshifts ($z\lesssim0.5$), which can be attributed to the extended structures of host galaxies \citep{HwangShen2020}. Our varstrometry-selected candidates generally have redshifts in the range of $z=1-3$. Even for a few low-redshift targets, they still have redshifts of $z\gtrsim0.5$. At the redshift range of most targets ($z=1-3$), the host galaxy emission is considered negligible in the Gaia bandpass. Furthermore, the optical DECaLS images show point-like morphology for the candidates that exhibit significant VLBA-Gaia offsets or multiple radio components. Based on these observations, we conclude that the Gaia positions are primarily based on the emission from the quasars, and the host galaxies are not the cause of the offsets.

\subsubsection{Reference coordinate system}
Differences in reference frames between the VLBA and Gaia images can introduce systematic biases when comparing the two datasets. The celestial reference frame of our VLBA images is tied to the International Celestial Reference System (ICRS), and most of our phase reference calibrators have positions based on the 3rd realization of the International Celestial Reference Frame \citep[ICRF3;][]{Charlot2020}. The Gaia celestial reference frame is officially aligned with ICRF3, with a large-scale systematic uncertainty ranging from 20 to 30 $\mu$as \citep{Mignard2018,Lindegren2018}. This systematic uncertainty of 20 - 30 $\mu$as is much smaller than the VLBA-Gaia positional offsets observed in the eight candidates.
Additionally, by analyzing the VLBA-Gaia positional offsets in the 21 VLBA-detected quasars, we did not observe any significant systematic error in positional offsets (see Figure \ref{fig:offset_gaia_vlba}). This lack of substantial systematic shifts supports a good alignment between the reference frames of Gaia and VLBA, indicating that the systematic uncertainties in the celestial reference frame do not account for the observed VLBA-Gaia offsets in the candidates.

\subsubsection{Chance superposition of an unrelated radio source}

The radio detection could be merely an unrelated radio source in the background or foreground.
To assess the likelihood of chance superposition, we computed the probability using number statistics from the Very Large Array Sky Survey \citep[VLASS,][]{Lacy2020}. Employing a 3$\sigma$ detection threshold of $\sim$0.5 mJy at 8.37 GHz and assuming a typical spectral index $\alpha=0.71$ \citep{Gordon2021}, sources with a flux density greater than 1 mJy should be observable at 3 GHz. Based on VLASS epoch 1 images, there are approximately 1.7$\times$10$^6$ radio sources with flux density exceeding 1 mJy across the entire 35,285 deg$^2$ footprint \citep{Gordon2020, Gordon2021}. Consequently, assuming a uniform distribution, the probability of encountering a random radio source with flux density $>$ 1 mJy within a 0\farcs5 radius is $\sim$3$\times$10$^{-6}$. Therefore, it is unlikely that our radio detection is a result of a chance superposition with an unrelated radio source. 

\subsection{Fraction of VLBA-Gaia offset quasars}

%\textbf{\color{red} It would be great to compare with the control sample of normal quasars also the actual distribution of the Gaia-VLBA offsets and the offset angle w.r.t. the jet angle (if possible), and not just the ``fraction''}

%\textbf{\color{blue} YCC: Done. I added the distribution of the Gaia-VLBA offsets. However, the two papers do not provide the values for the offset angle w.r.t. the jet angle.}

To assess whether varstrometry is successful in identifying candidate dual and offset quasars, we compare our sample with several quantitative studies focusing on normal single quasars that have both VLBA and Gaia matches. Several studies have \citet{Petrov2019} analyzed systematic offsets in the positions of 9,081 matched sources between Gaia DR2 and VLBI. 
They found that 9\% of sources exhibit offsets statistically significant at the 99\% confidence level. \citet{Makarov2019} investigated a sample of 3,413 extragalactic sources from ICRF3. After excluding potential contamination from double sources, confusion sources, extended sources, and sources with poor quality, they found that 20\% of the sources had VLBA-Gaia offsets exceeding the 3$\sigma$ significance level.

\begin{figure}
    \centering
    \includegraphics[width=\columnwidth]{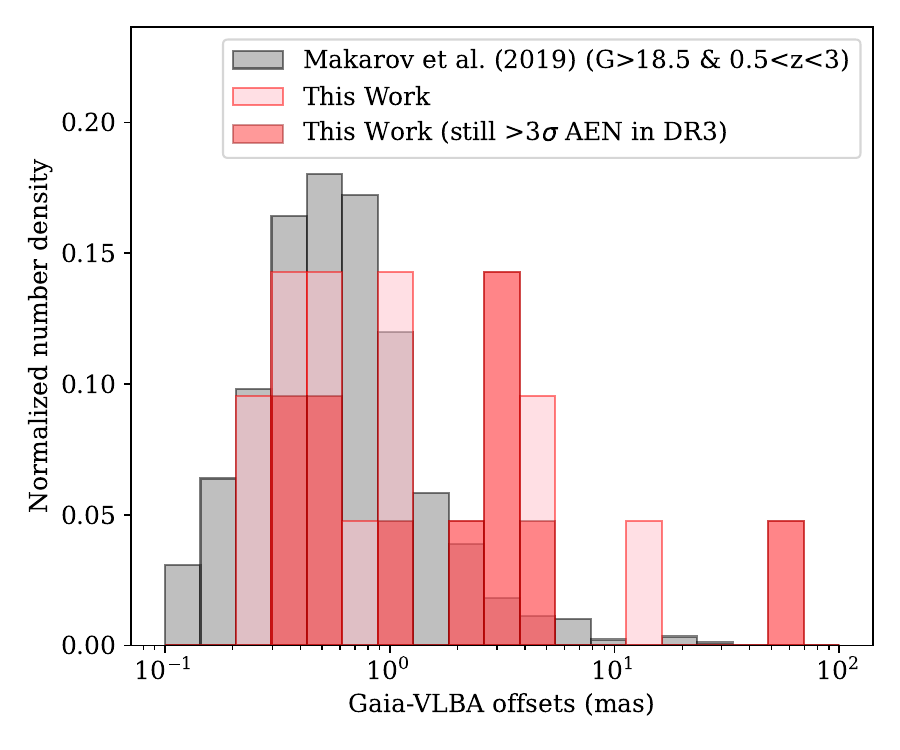}
    \caption{Distribution of VLBA-Gaia offsets for the 21 targets that have a detection in our sample (in red) compared to those for single normal quasars (in grey) in \citet{Makarov2019}. We apply the criteria of G $>$ 18.5 and 0.5 $<$ redshift $<$ 3 to the quasars from the literature to ensure comparability with our sample. We also show the subset of the 21 targets that still pass varstrometry selection using Gaia DR3 (in dark red). Our sample shows a notable excess in large VLBA-Gaia offsets.}
    \label{fig:offset_comparison}
\end{figure}

Our detection rate of 26$\pm$8\% (6/23, assuming 1$\sigma$ Poisson errors) is slightly higher but still broadly consistent ($<3\sigma$) with the rates reported in the literature for normal single quasars \citep{Petrov2019,Makarov2019}. Figure \ref{fig:offset_comparison} illustrates the distribution of VLBA-Gaia offsets for our sample compared to the offsets reported in \citet{Makarov2019}. To ensure comparability with our sample, the quasars from the literature are selected using the criteria : G $>$ 18.5 and $0.5 < z < 3$. Our varstrometry-selected sample includes a higher number of targets with significant VLBA-Gaia offsets (the p-values of two-sample Kolmogorov-Smirnov test is 0.02), indicating the potential of the varstrometry technique to discover additional targets with such offsets ($\gtrsim10 $ mas). If we only select the subset of the 21 targets that still pass varstrometry selection in Gaia DR3, the discrepancy between our sample and the control sample becomes even larger (p-values of 0.01). A larger sample with lower statistical uncertainty is needed to confirm this discrepancy definitively. Besides, the median flux density of the ICRF3 sources \citep[$\sim$0.13 Jy at S band,][]{Hunt2021} is one order of magnitude higher than those in our sample ($\sim$0.02 Jy at S band). Thus, the \citet{Makarov2019} sample, based on the ICRF3 sources, might contain a large portion of small-scale jets and be toward extremely radio-loud quasars. 

The varstrometry technique, theoretically, is more sensitive to scales of dozens or hundreds of mas than just a few mas, given the current accuracy of Gaia. According to the varstrometry assumption (Equations 3 and 5 in \citealt{HwangShen2020}), the typical astrometric signal for pairs with equal fluxes and 10\% flux variability is approximately 5\% of the pair separation. Therefore, for the typical Gaia astrometric excess noise of a few mas in our sample, the expected separations would fall within the range of dozens to hundreds of mas. Consequently, we believe that targets with offsets of around 10-100 mas could potentially be discovered using varstrometry, while those at a few mas scales are too small to be detected given the current astrometric accuracy of Gaia.

%Therefore, our detection rate is the combination of 
%For the larger We discover J0749+2255, which is a 0\farcs46 dual quasar and J1557$-$1236, which has a 0\farcs06 VLBA-Gaia offset. Discovering those two targets are a good demenstration of  
 
%we can not exclude the possibility that the offset in any of the five candidates indeed link to high astrometric noise in Gaia, especially for the targets having a large offset, like J1557$-$1236.

%\subsection{Gaia DR3}
%Gaia DR3 was released in June 2022 \citep{Brown2021}. The DR3 data provide better astrometric solutions by spanning a period of 34 months, compared to 22 months for the DR2 data \citep{Lindegren2021}. 
%We calculate the VLBA-Gaia offsets using new Gaia DR3 data and check any changes in the offsets compared with the values from DR2. The Gaia DR3 positions of all targets are consistent with the DR2 positions. Based on the same selection criteria, the five dual/offset AGN candidates still show significant offsets using Gaia DR3 data. All other targets remain consistent positions between VLBA and Gaia DR3. The unchanged results reaffirm that the significance of the VLBA-Gaia offsets seen in our sample, not originated from particular Gaia processing pipeline. 
%\subsection{Evidence for Dual/Offset Quasars?}

%\subsection{Small-scale Jets}

%Lensed quasars should have similar colors in the two \hst\ bands. 

%\subsection{Chance Superposition with Stars}

\section{Conclusions}\label{sec:conclusions}

We presented VLBA images and measurements of 23 radio-bright quasars selected by varstrometry, which have larger Gaia astrometric noise, aiming to search for candidate dual/off-nucleus quasars at pc scales. Out of the 23 quasars, 8 exhibit significant positional offsets between Gaia and VLBA or multiple radio components.

Out of the 8 candidates, 3 (J1051+2119, J1110+4817, and J1259+5140) exhibits multiple radio components. Based on the morphology and jet direction observed in single-band images, the radio emission is likely originating from small-scale jets, although confirmation through spectral index analysis using deep multiple-band observations is required.

The remaining 5 candidates with significant VLBA-Gaia offsets exhibit different characteristics. Three candidates (J0749+2255, J1044+2959, and J1110+3653) have compact radio cores with flat or inverted spectra, one candidate (J0108$-$0400) has extended radio emission with a steep spectrum, and one candidate (J1557+1236) has a large offset of 62.1 mas but lacks spectral index information due to only having a single-band image. Based on spectral indices and morphology, the VLBA-Gaia offset in J0108$-$0400 likely originates from small-scale radio jets. The offset in J0749+2255 could be attributed to blended optical emission in Gaia from a confirmed dual quasar on a larger scale of 0\farcs46 \citep{ChenYC2023}. J1044+2959 and J1110+3653 are possible candidate dual quasars at pc scales. The origin of the radio emission in J1557+1236 remains unknown due to weak detection and the lack of spectral index information. Deep multi-band follow-up imaging is necessary to confirm the properties of J1557+1236.

%\textbf{\color{red} Good start, but would be great to expand further on this and give examples, references, and/or more concrete details.} \textbf{\color{blue} YCC: Done. I added more details and references in this paragraph. }
In addition to the radio jet and dual/off-nucleus quasar scenarios, we explored other possible causes of the VLBA-Gaia offset. 
Optical jets at scales of hundreds of milli-arcseconds are considered a potential cause of VLBA-Gaia offsets \citep{Petrov2017,Petrov2019}.
However, examination of the optical spectra and images for the eight candidates did not find any evidence of optical jets contaminating the spectra or extended jets at arcsecond scales. While strong optical jets at arcsecond scales are unlikely to be responsible for the offset, the presence of faint sub-arcsecond optical jets cannot be ruled out.
Another potential factor is optical emission from host galaxies, which can affect Gaia's astrometric solutions. Most of our varstrometry-selected candidates are at high redshifts ($z=1-3$), so the host galaxy emission is negligible. 
Furthermore, differences in reference coordinate systems between VLBA and Gaia images can introduce systematic biases. The small systematic uncertainty of $20-30\mu$as \citep{Lindegren2018} and the absence of significant systematic shifts in the 21 VLBA-detected quasars, suggests that systematic uncertainties in the celestial reference frame is not a major contributor for the VLBA-Gaia offsets we observe.

We compare our varstrometry-selected sample and samples of normal single quasars from the literature \citep{Petrov2019, Makarov2019} to assess whether varstrometry can select a higher fraction of quasars with VLBA-Gaia offsets. 
We find that the fraction of quasars with significant ($>3\sigma$) offsets in our sample is slightly higher, but still broadly consistent with those in the literature, taking into account the uncertainty. Additionally, we examine the distribution of the VLBA-Gaia offsets and observe a greater number of large offsets in our varstrometry-selected sample, in comparison with normal single quasars, demonstrating the potential of the varstrometry technique to uncover more candidate dual or off-nucleus quasars.

\facilities{Gaia, VLBA}.

\software{AIPS, astropy \citep{Astropy2013,Astropy2018,Astropy2022}, numpy \citep{Numpy2011,Numpy2020}}.

\acknowledgments
We thank the anonymous referee for giving constructive comments. We thank Eric Greisen, Meri Stanley, and other observers for their help with our VLBA observations and data reduction.
This work is supported by the Heising-Simons Foundation and Research Corporation for Science Advancement, and NSF grants AST-2108162 and AST-2206499 (YCC, XL, YS). YCC and XL acknowledge support from the University of Illinois Campus Research Board. YCC acknowledges support by the government scholarship to study abroad from the ministry of education of Taiwan and support by the Illinois Survey Science Graduate Student Fellowship. YS acknowledges partial support from NSF grant AST-2009947. This research was supported in part by the National Science Foundation under PHY-1748958.
SBS gratefully acknowledges the support of a Sloan Fellowship, and the support of NSF under award \#1815664.
The National Radio Astronomy Observatory is a facility of the National Science Foundation operated under cooperative agreement by Associated Universities, Inc. 
This work made use of the Swinburne University of Technology software correlator, developed as part of the Australian Major National Research Facilities Programme and operated under licence \citep{Deller2011}.
This work has made use of data from the European Space Agency (ESA) mission Gaia (\url{https://www.cosmos.esa.int/gaia}), processed by the Gaia
Data Processing and Analysis Consortium (DPAC,
\url{https://www.cosmos.esa.int/web/gaia/dpac/consortium}). Funding for the DPAC
has been provided by national institutions, in particular the institutions
participating in the Gaia Multilateral Agreement.
%NLZ and HCH acknowledge support by the HST-SNAP-15900 grant administered by the STScI.

%Based on observations made with the NASA/ESA Hubble Space Telescope, obtained at the Space Telescope Science Institute, which is operated by the Association of Universities for Research in Astronomy, Inc., under NASA contract NAS 5-26555. These observations are associated with program number GO 15900.

\bibliography{ref}

\end{document}